\documentclass[11pt,a4paper]{article}
\pdfoutput=1

\usepackage{jheppub}
\usepackage[rgb]{xcolor}
\usepackage{color}
\usepackage{amsmath,amsfonts,amssymb,mathrsfs,graphicx,bbm,dsfont,booktabs,mathtools,braket,lmodern,tikz,slashed}

\DeclareMathAlphabet{\mathpzc}{OT1}{pzc}{m}{it}

\newcommand{\bea}{\begin{eqnarray}}
\newcommand{\eea}{\end{eqnarray}}
\def\be{\begin{equation}}
\def\ee{\end{equation}}
\newcommand{\bei}{\begin{itemize}}
\newcommand{\eei}{\end{itemize}}
\newcommand{\bee}{\begin{enumerate}}
\newcommand{\eee}{\end{enumerate}}

\def\ads{{\rm AdS}_5\times {\rm S}^5}

\def\ads{{\rm AdS}_5\times {\rm S}^5}

\def\am{{\rm am}}
\def\am0{{\rm am}_0}

\DeclareMathAlphabet{\mathsc}{T1}{lmr}{m}{scsl}

\expandafter\def\expandafter\bfseries\expandafter{\bfseries\ifmmode\else\boldmath\fi}
\expandafter\def\expandafter\mdseries\expandafter{\mdseries\ifmmode\else\unboldmath\fi}
\expandafter\def\expandafter\normalfont\expandafter{\normalfont\ifmmode\else\unboldmath\fi}

\def\agauge{{a}}

\definecolor{grey}{rgb}{0.4,0.4,0.5}
\definecolor{darkgreen}{rgb}{0,0.5,0}
\definecolor{darkred}{rgb}{0.6,0.0,0}
\definecolor{lightbrown}{rgb}{1,0.9,0.8}
\definecolor{brown}{rgb}{0.6,0.3,0.3}
\definecolor{darkblue}{rgb}{0,0,0.8}
\definecolor{darkmagenta}{rgb}{0.5,0,0.5}

\def\Anot{C}

\title{Double Wick rotating Green-Schwarz strings}

\author[a,1]{Gleb Arutyunov}
\note{Correspondent fellow at Steklov Mathematical Institute, Moscow.}
\author[b]{and Stijn J. van Tongeren}

\affiliation[a]{Institut f\"ur Theoretische Physik, Universit\"at Hamburg, Luruper Chaussee 149, 22761 Hamburg, Germany}
\affiliation[a]{Zentrum f\"ur Mathematische Physik, Universit\"at Hamburg, Bundesstrasse 55, 20146 Hamburg, Germany}
\affiliation[b]{Institut f\"ur Mathematik und Institut f\"ur Physik, Humboldt-Universit\"at zu Berlin, IRIS Geb\"aude, Zum Grossen Windkanal 6, 12489 Berlin, Germany}

\emailAdd{gleb.arutyunov@desy.de}
\emailAdd{svantongeren@physik.hu-berlin.de}

\abstract{Via an appropriate field redefinition of the fermions, we find a set of conditions under which light cone gauge fixed world sheet theories of strings on two different backgrounds are related by a double Wick rotation. These conditions take the form of a set of transformation laws for the background fields, complementing a set of transformation laws for the metric and B field we found previously with a set for the dilaton and RR fields, and are compatible with the supergravity equations of motion. Our results prove that at least to second order in fermions, the $\ads$ mirror model which plays an important role in the field of integrability in AdS/CFT, represents a string on `mirror $\ads$', the background that follows from our transformations. We discuss analogous solutions for $\rm{AdS}_3\times \rm{S}^3 \times \rm{T}^4$ and $\rm{AdS}_2\times \rm{S}^2 \times \rm{T}^6$. The main ingredient in our derivation is the light cone gauge fixed action for a string on an (almost) completely generic background, which we explicitly derive to second order in fermions.}

\begin{document}

\begin{flushright}\small{HU-EP-14/65\\HU-MATH-14/39\\ZMP-HH-14/26}\end{flushright}

\maketitle

\section{Introduction}

A double Wick rotation on the world sheet of a string is a transformation that has proven its use in the context of the AdS/CFT correspondence \cite{Maldacena:1997re}. Fixing a light cone gauge in the world sheet theory of a superstring on $\ads$ \cite{Metsaev:1998it} results in a model that transforms nontrivially under a double Wick rotation, and it is the thermodynamics of this different two dimensional quantum field theory \cite{Ambjorn:2005wa,Arutyunov:2007tc} that lies at the heart of the solution of the spectral problem in AdS/CFT using integrability \cite{Arutyunov:2009ga,Beisert:2010jr}. Namely, the spectrum of the string on $\ads$ can be computed by means of the thermodynamic Bethe ansatz for this so-called mirror model \cite{Arutyunov:2009zu,Arutyunov:2009ur,Bombardelli:2009ns,Gromov:2009bc}.\footnote{These equations have been used to get impressive finite coupling results (see e.g. \cite{Frolov:2010wt}), and have been simplified to the so-called quantum spectral curve \cite{Gromov:2013pga}, particularly impressive in the perturbative regime (see e.g. \cite{Marboe:2014gma}).}

We may wonder whether such double Wick rotated models have a more direct physical interpretation. In previous work we showed that (given mild restrictions) doing a double Wick rotation at the bosonic level is equivalent to a particular transformation of the metric and B field \cite{Arutyunov:2014cra}. Here we will establish the analogous result for the fermions, and find the transformation of (a combination of) the dilaton and Ramond-Ramond (RR) background fields such that the complete transformation is equivalent to a double Wick rotation of the Green-Schwarz action (to second order in fermions).

In order to find this transformation, we take a type IIB Green-Schwarz string to second order in fermions on an (almost) completely generic background with light cone isometries, and fix a light cone gauge. We then specify our restrictions on the metric and B field and get the set of fermionic terms that are to be analytically continued. As fermions have single derivative kinetic terms, their double Wick rotation is subtle and requires a complex field redefinition; fortunately this can be unambiguously fixed in the flat space limit. We then proceed by attempting to `undo' the double Wick rotation, knowing the transformation that accomplishes this at the bosonic level. Hence we take the general double Wick rotated action and rewrite it in terms of the transformed metric and B field, changing the light cone block of the metric as $g_{\rm{lc}}\rightarrow g_{\rm{lc}}/|\det g_{\rm{lc}}|$ and the sign of the (transverse) B field. As an immediate consistency check we find that all fermionic terms fixed by the metric and B field directly match the original action as they must. We then compare the double Wick rotated dilaton--RR terms (with transformed metric and B field) to the originals and read off how they should transform. For our considerations to apply straightforwardly, we find that similarly to how we assume our metric and B field not to mix light cone and transverse directions, the RR fields should not lead to such mixing either. This translates to the requirement that the RR fields should always have a single light cone index. The dilaton and allowed RR fields then need to transform such that the combinations $e^{\Phi} F^{(1/5)}_{a(bcde)}$ are invariant, while $e^\Phi F^{(3)}_{abc}$ picks up a sign, like $dB$. Of course these RR fields now also live in a different space, so that the unchanging nature of $e^\Phi F$ is only apparent. While we do not work out the details, the same analysis can be readily repeated for the type IIA string.

Then, by tracing the world sheet double Wick rotation back through the light cone gauge fixing procedure,\footnote{We would like to thank A. Tseytlin for this suggestion} we discuss how to interpret the transformation of the background fields from a target space perspective, as a combination of T dualities and target space analytic continuation (Wick rotation). Though the analytic continuation is somewhat subtle, this shows that our transformation is compatible with the supergravity equations of motion, precisely under our restrictions on the background fields.

Our main example is $\ads$ and its cousin `mirror $\ads$', whose supergravity solution we constructed in \cite{Arutyunov:2014cra}. The associated background fields match our transformations, thereby explicitly proving (to second order in the fermions) that the $\ads$ mirror model physically represents a string on mirror $\ads$. Interestingly, mirror $\ads$ inherits the integrability and supersymmetry of the $\ads$ light cone sigma model, despite mirror $\ads$ being a singular and nonsupersymmetric solution of supergravity \cite{Arutyunov:2014cra}. Our current results show that this supersymmetry is indeed realized directly on the canonical Green-Schwarz fermions of the background, albeit nonlinearly \cite{Arutyunov:2014cra}.  We briefly discuss similar mirror solutions for $\rm{AdS}_3\times \rm{S}^3 \times \rm{T}^4$ and $\rm{AdS}_2\times \rm{S}^2 \times \rm{T}^6$. While our restrictions on the background do not appear too drastic, spaces such as e.g. $\rm{AdS}_4 \times \mathbb{CP}^3$ do not fit them. In the outlook we discuss the apparent obstacles that arise when we relax our restrictions.

We should note that our work was ultimately inspired by the `mirror duality' \cite{Arutynov:2014ota} of the integrable deformation of the $\ads$ coset model of \cite{Delduc:2013qra,Delduc:2014kha}.\footnote{For further developments in this area see e.g. \cite{Arutyunov:2013ega,Hoare:2014pna,Kameyama:2014bua,Khouchen:2014kaa,Ahn:2014aqa,Arutyunov:2014cda,Banerjee:2014bca,Kameyama:2014vma,Hollowood:2014qma,Sfetsos:2014cea,Lunin:2014tsa,Hoare:2014oua,Matsumoto:2014ubv}.} Labeling the deformation by a parameter $\kappa$, an exact S-matrix approach to this model showed that the light cone theory at $1/\kappa$ is equivalent to the double Wick rotation of the theory at $\kappa$ \cite{Arutynov:2014ota}. This statement has a simple proof at the bosonic level using the metric and B field found in \cite{Arutyunov:2013ega}, and in the $\kappa\rightarrow0$ limit gives precisely (mirror) $\ads$ \cite{Arutyunov:2014cra}. The results of our paper should allow us to check this statement at the fermionic level, assuming this model corresponds to a string sigma model. However, to date the dilaton and RR fields of this model have not been found, and whether this and related models represent strings for generic values of $\kappa$ is an interesting open question. We will briefly come back to this in our conclusions.

In the next section we will begin our discussion by summarizing the transformation of the metric and B field that is equivalent to a double Wick rotation at the bosonic level. In section \ref{sec:GSferms} we introduce fermions on the world sheet, and present the light cone gauge fixed action derived in appendix \ref{app:lightconegaugefixing}. We then discuss the double Wick rotation of fermions, the matching of the metric and B field terms, and the transformations of and constraints on the dilaton and RR fields in section \ref{sec:fermmirortf}. Section \ref{sec:doubleWickBackgrounds} contains the discussion of our transformation from a target space perspective, along with the examples of $\ads$, $\rm{AdS}_3\times \rm{S}^3 \times \rm{T}^4$, and $\rm{AdS}_2\times \rm{S}^2 \times \rm{T}^6$ and their mirror partners. We finish in section \ref{sec:conclusions} by discussing open questions and generalizations of our work.

\section{Wick rotated bosons and changed geometry}
\label{sec:bosmirrortf}

To set the stage, let us recall the change of geometry that is equivalent to doing a double Wick rotation on the world sheet of a bosonic light cone string \cite{Arutyunov:2014cra}.

The action describing bosonic string propagation on a generic background is given by
\begin{equation}
\label{eq:bosstringaction}\nonumber
S \equiv -\tfrac{T}{2}\int{\rm d}\tau{\rm d}\sigma \mathcal{L} =-\tfrac{T}{2}\int{\rm d}\tau{\rm d}\sigma \, \left(g_{\mathsc{mn}}\, dx^\mathsc{m} dx^\mathsc{n} - B_{\mathsc{mn}}\, dx^\mathsc{m} \hspace{-3pt} \wedge dx^\mathsc{n}\right) \,,
\end{equation}
where $T$ is the (effective) string tension. We will consider $d$-dimensional backgrounds with coordinates $\{t,\phi,x^\mu\}$, packaging $t$ and $\phi$ in the (generalized) light cone combinations
\bea
x^+=(1-\agauge)t+\agauge\phi , \, \mbox{ and } x^{-}=\phi-t\, ,
\eea
where $\agauge$ is a free parameter.\footnote{At the level of gauge fixing, this parameter interpolates between the temporal gauge at $a=0$ and the conventional light cone gauge at $a=1/2$ \cite{Arutyunov:2006gs}. In \cite{Arutyunov:2014cra} we took $a=0$ for conciseness.} Further conventions are summarized in appendix \ref{app:Gammamatricesandkappafixing}. We take the metric to be of the block diagonal form
\begin{equation}
\label{eq:generalmetric}
g_{\mathsc{mn}} = \left(\begin{array}{ccc}g_{++} & g_{+-} & 0\\
g_{+-} & g_{--} & 0 \\
0 & 0 & g_{\mu\nu}\end{array}\right) \equiv \left(\begin{array}{cc} g_{\rm{lc}} & 0 \\
0 & g_{\mu\nu}\end{array}\right),
\end{equation}
with components depending only on the transverse coordinates $x^\mu$, and consider purely transverse B fields. We will address the question of relaxing these restrictions in section \ref{sec:conclusions}.

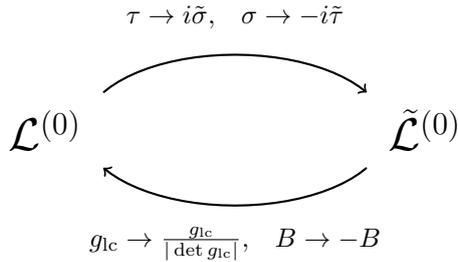
\begin{figure}
\begin{center}
\begin{tikzpicture}
\node at (-5/2,0) {{\LARGE $\mathcal{L}^{(0)}$}};
\node at (5/2,0) {{\LARGE $\tilde{\mathcal{L}}^{(0)}$}};
\node at (0,3/2) {\small $\tau \rightarrow i \tilde{\sigma},$ \hspace{3pt} $\sigma \rightarrow -i \tilde{\tau}$};
\node at (0,-3/2) {\small $g_{\rm{lc}}\rightarrow \frac{g_{\rm{lc}}}{|\det g_{\rm{lc}}|},$ \hspace{3pt} $B\rightarrow-B$};
\draw[black, thick,->, domain=-pi/3:pi/3] plot ({2*sin(\x r)},{cos(\x r)});
\draw[black, thick,<-, domain=-pi/3:pi/3] plot ({2*sin(\x r)},{-1*cos(\x r)});
\end{tikzpicture}
\end{center}
\caption{A double Wick rotation as a change of geometry. At the bosonic level a double Wick rotation of the light cone world sheet theory of a string is equivalent to a change of the metric and B field.}
\label{fig:bosmirrortf}
\end{figure}

As discussed in more detail in appendix \ref{app:lightconegaugefixing}, to fix a uniform light cone gauge we impose\footnote{We focus on the zero winding sector in cases where $\phi$ parametrizes a circle the string can wind around.}
\begin{equation}
x^+ = \tau \,, \hspace{20pt} p_- = 1\,.
\end{equation}
The resulting action takes the form
\begin{equation}
\label{eq:bosgfaction}
S^{(0)}=-T\int{\rm d}\tau{\rm d}\sigma \, \left( \sqrt{Y^{(0)}} + g^{+-}/g^{++} -\dot{x}^{\mu}x'^{\nu}B_{\mu\nu}\right)\, ,
\end{equation}
where
\begin{equation}
\label{eq:Y0def}
Y^{(0)} = (\dot{x}^\mu x^\prime_\mu)^2 - \frac{A\Anot}{g^{++} g_{--}}
\end{equation}
with
\begin{equation}
\label{eq:Adefinition}
\begin{aligned}
A =&\, 1+g_{--} x^{\prime \mu}x^\prime_\mu,\\
\Anot =&\, 1+g^{++} \dot{x}^\nu \dot{x}_\nu,
\end{aligned}
\end{equation}
and dots and primes refer to derivatives with respect to time $(\tau)$ respectively space $(\sigma)$. Note that we work in units that put spatial and temporal derivatives on the same footing. Now we readily see that a double Wick rotation of the world sheet coordinates
\begin{equation}
\label{eq:mirrortf}
\tau  \rightarrow i \tilde{\sigma}, \hspace{20pt} \sigma \rightarrow - i \tilde{\tau},
\end{equation}
gives an action of the same form, with $g_{--}$ exchanged for $- g^{++}$ while leaving $g^{+-}/g^{++}$ unchanged,\footnote{Since $g^{+-}/g^{++} = -g_{+-}/g_{--}$, this is equivalent to exchanging $g^{+-}$ for $g_{+-}$.} and with a change of sign on $B$. This change of the metric is nothing but
\begin{equation}
\label{eq:metricchange}
g_{\rm{lc}}\rightarrow \epsilon g_{\rm{lc}}^{-1} \epsilon = \frac{g_{\rm{lc}}}{|\det g_{\rm{lc}}|},
\end{equation}
which applies in any coordinate system in the light cone subspace. In particular it is simply $g_{tt} \leftrightarrow g^{\phi\phi}$ when the original metric is diagonal in $t$ and $\phi$.\footnote{We define $ds^2 = - g_{tt} dt^2\,\, (+ 2g_{t\phi} dt d\phi) + g_{\phi\phi} d\phi^2 + g_{\mu\nu} dx^\mu dx^\nu$.} Put differently, the action is formally left invariant under a double Wick rotation combined with the transformations \eqref{eq:metricchange} and $B \rightarrow -B$ on the background. We have illustrated this idea in figure \ref{fig:bosmirrortf}. Keep in mind that this combined transformation interchanges $A$ and $\Anot$ and leaves $Y^{(0)}$ invariant.

At this stage it is not clear that the mirror space associated to a string background represents a string background itself. We will come back to this point in detail later, for now let us simply assume there are examples where this is the case. We would then like to see that the fermions of such strings behave appropriately under this transformation as well. Let us therefore add fermions to our string, and fix a light cone gauge again, so that we can determine the required transformation properties of the dilaton and the RR fields.

\section{Green-Schwarz fermions}
\label{sec:GSferms}

As we are interested in string backgrounds with RR fields, we need to use the Green-Schwarz formalism. Though a fully explicit construction is not known, the Lagrangian describing propagation of a Green-Schwarz superstring in an arbitrary background can be constructed second order by second order in the fermions, explicitly done up to fourth order in \cite{Wulff:2013kga}.\footnote{To second order this Lagrangian was first derived in \cite{Tseytlin:1996hs,Cvetic:1999zs}.} We will limit ourselves to second order, which will provide a nontrivial test of our ideas. As $\ads$ is one of our spaces of interest, we will focus on the type IIB string. The Lagrangian (density) at quadratic order in the fermions is given by \cite{Wulff:2013kga}
\begin{equation}
\mathcal{L}^{(2)}_f = i\left( \star e^a \bar{\theta} \Gamma_a \mathcal{D} \theta - e^a \bar{\theta} \Gamma_a \sigma_3 \mathcal{D} \theta\right)\,,
\end{equation}
where $\theta = (\theta_1,\theta_2)$ is a doublet of sixteen component Majorana-Weyl spinors, with $\sigma_3$ and the $\epsilon \equiv i\sigma_2$ term in $\mathcal{D}$ (see below) acting in this two dimensional space. $\mathcal{D}$ is given by
\begin{equation}
\mathcal{D} = d -\tfrac{1}{4} \slashed{\omega} + \tfrac{1}{8} e^a H_{abc} \Gamma^{bc} \sigma_3 +\tfrac{1}{8} e^b \mathcal{S} \Gamma_b,
\end{equation}
where $\omega$ is the spin connection, $e$ the (one form) vielbein, and $H=dB$ is the Neveu-Schwarz--Neveu-Schwarz (NSNS) three form. Slashes denote contraction with the appropriate product of $\Gamma$ matrices, $\Gamma_{a\ldots m}$. $\mathcal{S}$ contains the dilaton and RR fields
\begin{equation}
\label{eq:Sformdefinition}
\mathcal{S} = - e^{\Phi}(\epsilon \slashed{F}^{(1)} + \tfrac{1}{3!} \sigma_1 \slashed{F}^{(3)} + \tfrac{1}{2\cdot 5!}\epsilon \slashed{F}^{(5)}),
\end{equation}
with the $F^{(n)}$ denoting the $n$ form RR fields, and $\Phi$ the dilaton. In components this reads
\begin{equation}
\mathcal{L}^{(2)}_f = i \gamma^{\alpha\beta} \partial_\alpha x^\mathsc{m} \bar{\theta} \Gamma_\mathsc{m} \mathcal{D}_\beta \theta - i \epsilon^{\alpha\beta} \partial_\alpha x^\mathsc{m} \bar{\theta} \Gamma_\mathsc{m} \sigma_3 \mathcal{D}_\beta \theta\,,
\end{equation}
with
\begin{equation}
\mathcal{D}_\beta = \partial_\beta -\tfrac{1}{4} \slashed{\omega}_\mathsc{m} \partial_\beta x^\mathsc{m}+ \tfrac{1}{8} \partial_\beta x^{\mathsc{m}} H_{\mathsc{mnp}}\Gamma^\mathsc{np}\sigma_3 + \tfrac{1}{8} \mathcal{S} \Gamma_\mathsc{m} \partial_\beta x^\mathsc{m} \,.
\end{equation}
We prefer to write out the Lagrangian as
\begin{equation}
\label{eq:Lagrangian}
\begin{aligned}
\mathcal{L} = \,&\gamma^{\alpha\beta} \hat{g}_{\mathsc{m}\mathsc{n}} \partial_\alpha x^\mathsc{m} \partial_\beta x^\mathsc{n} - \epsilon^{\alpha\beta} \hat{B}_{\mathsc{m}\mathsc{n}} \partial_\alpha x^\mathsc{m} \partial_\beta x^\mathsc{n} \\
& + i\gamma^{\alpha \beta} \partial_\alpha x^\mathsc{m} \bar{\theta} \Gamma_\mathsc{m} \partial_\beta\theta - i\epsilon^{\alpha \beta} \partial_\alpha x^\mathsc{m} \bar{\theta} \Gamma_\mathsc{m} \sigma_3 \partial_\beta \theta \,,
\end{aligned}
\end{equation}
where the hats indicate these terms are not purely the bosonic metric and B field, i.e.
\begin{equation}
\label{eq:Bhatandghatdef}
\begin{aligned}
\hat{g}_{\mathsc{m}\mathsc{n}} &\, = g_{\mathsc{m}\mathsc{n}} - \tfrac{i}{4} \bar{\theta} \Gamma_{(\mathsc{m}} \slashed{\omega}_{\mathsc{n})} \theta + \tfrac{i}{8} \bar{\theta} \Gamma_{(\mathsc{m}}H_{\mathsc{n})\mathsc{pq}}\Gamma^{\mathsc{pq}} \sigma_3 \theta + \tfrac{i}{8} \bar{\theta} \Gamma_{(\mathsc{m}} \mathcal{S} \Gamma_{\mathsc{n})} \theta,
 \\
\hat{B}_{\mathsc{m}\mathsc{n}} &\, = B_{\mathsc{m}\mathsc{n}} - \tfrac{i}{4} \bar{\theta} \Gamma_{[\mathsc{m}} \slashed{\omega}_{\mathsc{n}]}\sigma_3 \theta  + \tfrac{i}{8} \bar{\theta} \Gamma_{[\mathsc{m}}H_{\mathsc{n}]\mathsc{pq}} \Gamma^{\mathsc{pq}}\theta + \tfrac{i}{8} \bar{\theta} \sigma_3 \Gamma_{[\mathsc{m}} \mathcal{S} \Gamma_{\mathsc{n}]} \theta,
\end{aligned}
\end{equation}
with round and rectangular brackets denoting symmetrization and antisymmetrization respectively, defined with the usual factor of $1/n!$. In this notation we will treat $\hat{g}$ as if it really were the metric, so it can be used to raise and lower indices at intermediate stages of computation.\footnote{The price of this hopefully unambiguous notation is having to keep track of extra signs in fermionic terms in the inverse `metric'. Note that while not explicitly done in this paper, this notation is very useful for example when fixing a light cone gauge in a Hamiltonian setting.} When light cone gauge fixing, the fermions that contribute to terms in $\hat{g}$ and $\hat{B}$ that are already nonzero at the bosonic level formally go along for the (bosonic) ride, while possible extra nonzero terms need only be kept at a linearized level (quadratic in fermions). Of course we still need to add the manifestly fermionic terms in eq. \eqref{eq:Lagrangian} to the derivation. The only assumption we will make at this stage is that the fermions do not generate a nonzero $\hat{B}_{+-}$, which will turn out to follow from our restrictions on the metric and B field and the restrictions on the RR fields we will find later.

\subsection*{Light cone action}

In appendix \ref{app:lightconegaugefixing} we fix a light cone gauge for a string on a completely generic background with light cone isometries and $B_{+-}=0$, including fermions to second order. For the type of backgrounds we are considering, this gives the gauge fixed action
\begin{equation}
\label{eq:gf-lagr-ferm}
S = S^{(0,2)} -T\int{\rm d}\tau{\rm d}\sigma \left( \mathcal{L}_{a}^{(2)} + \mathcal{L}_{b}^{(2)}\right)\,,
\end{equation}
where  $S^{(0,2)}$ is the bosonic gauge fixed action \eqref{eq:bosgfaction} with metric and B field replaced by (the relevant parts of) $\hat{g}$ and $\hat{B}$. $\mathcal{L}_{a}^{(2)}$ and $\mathcal{L}_{b}^{(2)}$ are given by
\begin{align}
\label{eq:Lfa}
\mathcal{L}_{a}^{(2)} = \frac{1}{2\sqrt{Y^{(0)}}}&\frac{1}{g_{--}g^{++}}\Big[(2 \hat{g}^{+\mu} \dot{x}_{\mu} - i \bar{\theta} \Gamma^+ \dot{\theta})A+(2\hat{B}_{-\mu} x^{\prime \mu} + i \bar{\theta} \Gamma_- \sigma_3 \theta^\prime)\Anot\\
 \nonumber & - \left(g_{--}(2\hat{g}^{+\mu} x^{\prime}_{\mu} - i \bar{\theta} \Gamma^+ \theta^\prime) + g^{++}(2\hat{B}_{-\mu} \dot{x}^\mu + i \bar{\theta} \Gamma_- \sigma_3 \dot{\theta})\right)\dot{x}_\mu x^{\prime\mu}\Big]
\end{align}
and
\begin{align}
\label{eq:Lfb}
\mathcal{L}_{b}^{(2)} = & \, -\frac{1}{g_{--}}\left(\frac{i}{2} \bar{\theta} \Gamma_- \dot{\theta} +\hat{g}_{-\mu} \dot{x}^\mu\right) - \frac{1}{g^{++}} \left( \frac{i}{2} \bar{\theta} \Gamma^+ \sigma_3 \theta^\prime + \hat{B}^{+\mu} x_{\mu}^\prime\right)
\end{align}
which are the formally new terms introduced by the fermions. $A$ and $\Anot$ are those of eqs. \eqref{eq:Adefinition}. In this action we have fixed a $\kappa$ symmetry gauge by taking
\begin{equation}
(\Gamma^0 + \Gamma^p) \theta = 0,
\end{equation}
where $\Gamma^0$ and $\Gamma^p$ are the flat space cousins of $\Gamma^t$ and $\Gamma^\phi$, see appendix \ref{app:Gammamatricesandkappafixing}. This means we essentially dropped terms of the form $\bar{\theta} \Gamma^\mu \theta$ in $\Gamma$ matrix structure; appendix \ref{app:lightconegaugefixing} contains the full action before imposing a $\kappa$ symmetry gauge choice.

\section{Wick rotated fermions and changed background fields}
\label{sec:fermmirortf}

In section \ref{sec:bosmirrortf} we recalled how at the bosonic level a double Wick rotation on the world sheet is actually equivalent to a change of background. We would now like to investigate this at the fermionic level. To do so, we will proceed as before, carefully considering a double Wick rotation of the general action \eqref{eq:gf-lagr-ferm}, and attempting to reinterpret the result as an action of the same form.

The story is slightly more involved for fermions than for bosons, due to their single derivative kinetic terms. Already for the massless free Dirac Lagrangian it is clear that a double Wick rotation does not result in a real Lagrangian if we keep the conjugation properties of the fermions fixed. Hence we need to accompany a double Wick rotation by a change of reality condition, or equivalently, a complex field redefinition.\footnote{In the context of the $\ads$ coset sigma model this was noted explicitly in \cite{Arutyunov:2007tc}, here we will be considering a transformation of the canonical Green-Schwarz fermions in a general setting.} We will fix this field redefinition by considering a Green-Schwarz string in flat space, which we will then use to find the double Wick rotated action on a general background.

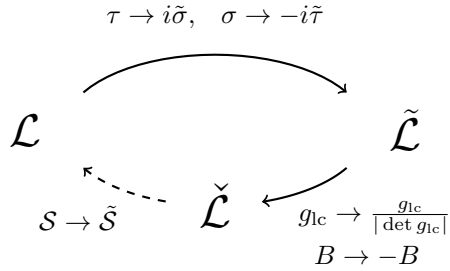
\begin{figure}
\begin{center}
\begin{tikzpicture}
\node at (-5/2,0) {{\LARGE $\mathcal{L}$}};
\node at (5/2,0) {{\LARGE $\tilde{\mathcal{L}}$}};
\node at (0,-1) {{\LARGE $\check{\mathcal{L}}$}};
\node at (0,3/2) {\small $\tau \rightarrow i \tilde{\sigma},$ \hspace{3pt} $\sigma \rightarrow -i \tilde{\tau}$};
\node at (2.1,-7/6) {\small $g_{\rm{lc}}\rightarrow \frac{g_{\rm{lc}}}{|\det g_{\rm{lc}}|}$ };
\node at (2,-10/6) {\small $B\rightarrow-B$};
\node at (-1.8,-7/6) {\small $\mathcal{S} \rightarrow \tilde{\mathcal{S}}$};
\draw[black, thick,->, domain=-pi/3:pi/3] plot ({2*sin(\x r)},{cos(\x r)});
\draw[black, thick,<-, domain=pi/10:pi/3] plot ({2*sin(\x r)},{-1*cos(\x r)});
\draw[black, dashed, thick,<-, domain=-pi/3:-pi/10] plot ({2*sin(\x r)},{-1*cos(\x r)});
\end{tikzpicture}
\end{center}
\caption{A double Wick rotation as a change of background fields? Combining a double Wick rotation with a change of the metric and B field and insisting the result agrees with the original action should fix the transformation $\mathcal{S} \rightarrow \tilde{\mathcal{S}}$ of the dilaton and RR fields. The intermediate $\check{\mathcal{L}}$ has no obvious physical interpretation.}
\label{fig:temptotalmirrortf}
\end{figure}

To reinterpret the result as an action of the form \eqref{eq:gf-lagr-ferm} again, we will attempt to complete the bosonic diagram of figure \ref{fig:bosmirrortf} to the one of figure \ref{fig:temptotalmirrortf}, finding the total transformation of the background fields that `undoes' the double Wick rotation.

\subsection{Double Wick rotated Green-Schwarz fermions}

To determine the precise transformation of our fermions under a double Wick rotation, we will consider a Green-Schwarz string in flat space. The light cone Lagrangian for such a string is given by\footnote{This can be obtained from our general action in flat space, taking the limit $a\rightarrow 1/2$ from below. Note that in flat space with $a=1/2$, $\Gamma^+= G^+$, where $G^+$ is the matrix involved in $\kappa$ symmetry gauge fixing as discussed in appendix \ref{app:Gammamatricesandkappafixing}.}
\begin{equation}
\label{eq:flatspaceL}
\mathcal{L} = \frac{T}{2}\left(\dot{x}^\mu \dot{x}_\mu-x^{\prime\mu} x^\prime_\mu + i\bar{\theta} \Gamma^- \dot{\theta} + i\bar{\theta} \Gamma^- \sigma_3 \theta^\prime \right).
\end{equation}
Now let us carefully do a double Wick rotation
\begin{equation}
\begin{aligned}
x(\tau,\sigma) &\rightarrow x(i \tilde{\sigma},-i\tilde{\tau}) \equiv y(\tilde{\tau},\tilde{\sigma})\\
\theta(\tau,\sigma) &\rightarrow \theta(i \tilde{\sigma},-i\tilde{\tau}) \equiv \tilde{\theta}(\tilde{\tau},\tilde{\sigma})
\end{aligned}
\end{equation}
so that we have
\begin{equation}
\begin{aligned}
x^\prime &= i \dot{y},\hspace{20pt}\dot{x} = -i y^\prime,\\
\theta^\prime &= i \dot{\tilde{\theta}},\hspace{20pt}\dot{\theta} = -i \tilde{\theta}^\prime,
\end{aligned}
\end{equation}
where dots and primes refer to derivatives in the first and second arguments. The resulting action then takes the form
\begin{equation}
\label{eq:flatspaceLnaivedoublewick}
\mathcal{L} = \frac{T}{2}\left(\dot{y}^\mu \dot{y}_\mu-y^{\prime\mu} y^\prime_\mu + \bar{\tilde{\theta}} \Gamma^- \tilde{\theta}^\prime - \bar{\tilde{\theta}} \Gamma^- \sigma_3 \dot{\tilde{\theta}} \right).
\end{equation}
We see that the bosonic part of this Lagrangian is formally the same as that of the original Lagrangian \eqref{eq:flatspaceL}, and we can assume $y$ to have the reality properties of $x$; typically they are not even distinguished. If we however assume $\tilde{\theta}(\tilde{\tau},\tilde{\sigma})$ to have the reality properties of $\theta(\tau,\sigma)$, this action will not be real when the original Lagrangian is. To fix this we should change the reality properties of $\tilde{\theta}$. Equivalently, we would like to consider a (constant, complex) field redefinition
\begin{equation}
\tilde{\theta} = M \eta,
\end{equation}
where the action is real when written in terms of $\eta$ with conventional reality properties. A priori it is not clear how to fix $M$ exactly, but we have an additional physical requirement to impose.

As a light cone string in flat space is a Lorentz invariant model (manifest in the NSR description), a double Wick rotation should leave all physical properties of the model invariant. The simplest way to realize this is to insist that the form of the Lagrangian is invariant under a double Wick rotation. Comparing eqs. \eqref{eq:flatspaceL} and \eqref{eq:flatspaceLnaivedoublewick}, we see that this is the case if
\begin{equation}
\bar{\eta} M^t \Gamma^- M \eta^\prime = i \bar{\eta} \Gamma^- \sigma_3 \eta^\prime,
\end{equation}
and
\begin{equation}
\bar{\eta} M^t \Gamma^- \sigma_3 M \dot{\eta} = -i \bar{\eta} \Gamma^- \dot{\eta}.
\end{equation}
Not having or wanting to touch the $\Gamma$ matrices we need
\begin{equation}
\label{eq:Mreqs}
M^t M = i \sigma_3\, \hspace{20pt} M^t \sigma_3 M = - i 1.
\end{equation}
This fixes
\begin{equation}
M = \left(\begin{array}{cc} 0 & b \\ c & 0 \end{array}\right),
\end{equation}
with $c^2 = - b^2 = i$, mixing the two Majorana-Weyl spinors. The overal sign of $M$ is clearly inconsequential, but this still leaves the choice of $bc=\pm1$. Later we will see that this choice is actually inconsequential as well. In summary, a double Wick rotation can be implemented by the replacements\footnote{A quick investigation based on the results in appendix \ref{app:lightconegaugefixing} indicates that the fermions set to zero in our $\kappa$ symmetry gauge can also be appropriately continued (though apparently with $M=1$) and fit in our story below. Note that different transformations are allowed as our $\kappa$ symmetry gauge condition clearly separates the fermions in two classes; they are even to be expected since this second class of fermions does not have canonical kinetic terms. Interestingly, without fixing a $\kappa$ symmetry gauge, the (light cone) $\kappa$ symmetry of a string will become the $\kappa$ symmetry of its mirror version under a double Wick rotation. To avoid unnecessary technical details we will not consider these extra fermions further.}
\begin{equation}
\begin{aligned}
x^\prime &\rightarrow i \dot{x},\hspace{20pt}\dot{x} \rightarrow -i x^\prime,\\
\theta \rightarrow  M \eta,&\hspace{20pt} \theta^\prime \rightarrow i M \dot{\eta},\hspace{20pt} \dot{\theta} \rightarrow -i M\eta^\prime,
\end{aligned}
\end{equation}
where we have stopped carefully distinguishing the bosons.

\subsection{The transformed Lagrangian}

Looking at figure \ref{fig:temptotalmirrortf} it is clear that we are primarily interested in $\check{\mathcal{L}}$, the result of transforming $\mathcal{L}$ by combining a double Wick rotation with a change of the metric and B field so that the bosonic part of the action is left invariant. This change comes with strings attached however, since it appears hard to give a target (or flat) space interpretation to the result of transforming the metric in a target space $\Gamma$ matrix; for a diagonal metric, $\Gamma^t = \sqrt{g^{tt}} \Gamma^0$ becomes $\sqrt{g_{\phi\phi}} \Gamma^0$ for example. This can be fixed by the automorphism of our Clifford algebra taking
\begin{equation}
\label{eq:automorphism}
\Gamma^{0/p} \rightarrow i \Gamma^{p/0},
\end{equation}
where $p$ is the other flat light cone direction, associated to $\phi$. This automorphism is compatible with our $\kappa$ symmetry gauge fixing, and is equivalent to a unitary change of basis by $e^{-i \frac{\pi}{4} \Gamma^0 \Gamma^p}$ on the fermions.\footnote{Of course, in our $\kappa$ symmetry gauge this is just multiplication by $e^{-i \frac{\pi}{4}}$, and actually does nothing to our expressions. This is not a particularly useful point of view however.} Combining the change of metric with this automorphism we get\footnote{Restricting ourselves to the 2d light cone block, starting with the original vielbein $e$ satisfying $e^t \eta e = g$ we can construct a `mirror' vielbein as $\tilde{e} = \epsilon g^{-1} e \sigma_1$, since the mirror metric is just $\epsilon g^{-1} \epsilon$. This means we have $g^{-1} e \eta \Gamma = \epsilon \,\tilde{e}\, \sigma_1 \eta \Gamma$, i.e. $e^\pm_a \Gamma^a = \mp\, \tilde{e}_{\,\mp}^{\,a} (\sigma_1)_{a}^{\,\,b} \eta_{bc} \Gamma^c$. Combined with the transformation $\Gamma \rightarrow i \sigma_1 \Gamma$ we get the desired result.}
\begin{equation}
\label{eq:gammapmtf}
\left(\begin{array}{c} \Gamma^{\pm}\\ \Gamma_{\pm}\end{array}\right) \rightarrow\left(\begin{array}{c} \check{\Gamma}^{\pm}\\ \check{\Gamma}_{\pm}\end{array}\right) \rightarrow \pm i \left(\begin{array}{c} \Gamma_{\mp}\\ -\Gamma^{\mp}\end{array}\right),
\end{equation}
where the rightmost $\Gamma$ matrices now have proper target space indices with respect to the changed metric. The factors of $i$ will disappear again since thanks to our $\kappa$ symmetry gauge we also have
\begin{equation}
\bar{\eta} \rightarrow i \bar{\eta}.
\end{equation}

Taking the action \eqref{eq:gf-lagr-ferm} and doing the above double Wick rotation, changing the metric and B field and implementing this automorphism we get
\begin{equation}
S = \check{S}^{(0,2)} -T\int{\rm d}\tilde{\tau}{\rm d}\tilde{\sigma} \left( \check{\mathcal{L}}_{a}^{(2)} + \check{\mathcal{L}}_{b}^{(2)}\right)\,,
\end{equation}
with
\begin{align}
\check{\mathcal{L}}_{a}^{(2)} = \frac{1}{2\sqrt{Y^{(0)}}}&\frac{1}{g_{--}g^{++}}\Big[(i 2 \check{\hat{B}}_{-\mu} \dot{x}^{\mu} - i \bar{\eta} \Gamma^+ \dot{\eta})A + (-i 2 \check{\hat{g}}^{+\mu} x^{\prime}_{\mu} + i \bar{\eta} \Gamma_- \sigma_3 \eta^\prime)\Anot \\
 \nonumber & - \left(g_{--}(i 2 \check{\hat{B}}_{-\mu} x^{\prime \mu} - i \bar{\eta} \Gamma^+ \eta^\prime) + g^{++}(-i 2 \check{\hat{g}}^{+\mu}  \dot{x}_\mu + i \bar{\eta} \Gamma_- \sigma_3 \dot{\eta})\right)\dot{x}_\mu x^{\prime\mu}\Big]
\end{align}
and
\begin{align}
\check{\mathcal{L}}_{b}^{(2)} = & \, -\frac{1}{g_{--}}\left(\frac{i}{2} \bar{\eta} \Gamma_- \dot{\eta} - i \check{\hat{B}}^{+\mu} \dot{x}_\mu\right) - \frac{1}{g^{++}} \left( \frac{i}{2} \bar{\eta} \Gamma^+ \sigma_3 \eta^\prime +i \check{\hat{g}}_{-\mu} x^{\prime \mu}\right).
\end{align}
Remaining checks denote transformed quantities we have temporarily left implicit. Comparing this to the general action \eqref{eq:gf-lagr-ferm}, we see that the manifestly fermionic terms match perfectly with $\eta =\theta$. From the remaining terms we read off that we need to have
\begin{equation}
\label{eq:tfreqs1}
\begin{aligned}
\check{\hat{g}}^{+\mu} &= i \hat{B}_{-}^{\,\,\,\,\mu}, & \check{\hat{B}}_{-}^{\,\,\,\,\mu} &= -i \hat{g}^{+\mu}, \\
\check{\hat{g}}_{-\mu} &= -i \hat{B}^{+}_{\,\,\,\,\mu}, & \check{\hat{B}}^{+}_{\,\,\,\,\mu} &= i \hat{g}_{-\mu}.
\end{aligned}
\end{equation}
We also have to check that the fermionic terms in $\check{S}^{(0,2)}$ behave as the metric and B field, i.e.
\begin{equation}
\label{eq:tfreqs2}
\begin{aligned}
\check{\hat{g}}_{--} &= -\hat{g}^{++}, & \check{\hat{g}}^{++} &= -\hat{g}_{--},\\
\check{\hat{g}}_{+-} &= \hat{g}^{+-}, & \check{\hat{g}}^{+-}& = \hat{g}_{+-},\\
\check{\hat{g}}_{\mu\nu} &= \hat{g}_{\mu\nu},& \check{\hat{B}}_{\mu\nu} &= -\hat{B}_{\mu\nu}.\\
\end{aligned}
\end{equation}
Finally, in order not to leave our present framework, the assumption $\hat{B}_{+-} =0$ needs to be preserved. We will begin by checking that the fermionic terms in $\hat{g}$ and $\hat{B}$ determined by the metric and B field actually behave as they should.

\subsection{Fermions with metric and B field}

The fermionic terms in $\hat{g}$ and $\hat{B}$ that have only the metric or B field in them are $\bar{\theta} \Gamma_\mathsc{m} \slashed{\omega}_\mathsc{n} \theta$ and $\bar{\theta} \Gamma_\mathsc{m} H_{\mathsc{npq}} \Gamma^{\mathsc{pq}} \sigma_3 \theta$, with an extra $\sigma_3$ inserted for $\hat{B}$. Our $\kappa$ symmetry gauge choice together with our restrictions on the metric and B field imply that many of these terms vanish. Using the block notation of eq. \eqref{eq:generalmetric} we have
\begin{equation}
\bar{\theta} \Gamma_\mathsc{m} \slashed{\omega}_\mathsc{n} \theta =
\renewcommand{\arraystretch}{0.8}
{
\left(
\begin{array}{cc}
0 & {  \bullet} \\
{  \bullet} & 0
\end{array}
\right)_{\mathsc{mn}}},
\end{equation}
where bullet points denote asymmetric (generically) nonzero contributions. In other words the spin connection term in $\hat{g}$ and $\hat{B}$ only contributes to $\hat{g}_{\pm \mu}$ and $\hat{B}_{\pm \mu}$, which by assumption have no bosonic term. $H$ contributes similarly
\begin{equation}
\bar{\theta} \Gamma_\mathsc{m}H_{\mathsc{npq}} \Gamma^{\mathsc{pq}} \theta =
\renewcommand{\arraystretch}{0.8}
{
\left(
\begin{array}{cc}
0 & {  \bullet} \\
{  \bullet} & 0
\end{array}
\right)_{\mathsc{mn}}}.
\end{equation}
These terms therefore only contribute to the relations \eqref{eq:tfreqs1}.

The first of eqs. \eqref{eq:tfreqs1} requires the spin connection terms to transform as
\begin{equation}
\label{eq:spinconfermtf}
\bar{\theta}\Gamma^{(+} \slashed{\omega}^{\mu)}\theta \rightarrow -i \bar{\theta} \Gamma_{[-} \slashed{\omega}^{\mu]}\sigma_3\theta
\end{equation}
under the full set of transformations used to arrive at $\check{\mathcal{L}}$, since the fermionic contribution to the inverse of the `metric' $\hat{g}$ has raised indices but an opposite sign. Now under these transformations the light cone components of the spin connection change as
\begin{equation}
\left(\begin{array}{c}\slashed{\omega}^\pm\\ \slashed{\omega}_\pm
\end{array}\right)\rightarrow \mp  i \left(\begin{array}{c}\slashed{\omega}_{\mp}\\
-\slashed{\omega}^{\mp}
\end{array}\right),
\end{equation}
while $\slashed{\omega}_\mu$ is unaffected, see appendix \ref{app:spinconnection} for details. Combining this with all other transformations we get
\begin{equation}
\nonumber
\bar{\theta}\left(\Gamma^{+} \slashed{\omega}^{\mu}+\Gamma^{\mu} \slashed{\omega}^{+}\right)\theta \rightarrow (i\bar{\eta})\left((i \Gamma_{-}) \slashed{\omega}^{\mu}+\Gamma^{\mu} (-i\slashed{\omega}_{-})\right)(i\sigma_3) \eta = - i \bar{\eta}\left(\Gamma_{-} \slashed{\omega}^{\mu}-\Gamma^{\mu} \slashed{\omega}_{-}\right)\sigma_3 \eta,
\end{equation}
nicely matching eq. \eqref{eq:spinconfermtf} upon identifying $\eta = \theta$. The rest of eqs. \eqref{eq:tfreqs1} works analogously, the relative sign arising cf. eqs. \eqref{eq:Mreqs}.

The $H$ contribution to these relations has an extra $\sigma_3$ but otherwise proceeds similarly. The last of eqs. \eqref{eq:tfreqs1} for example requires
\begin{equation}
\nonumber
\bar{\theta}\Gamma^{[+}H_{\mu]\nu\rho} \Gamma^{\nu\rho} \theta \rightarrow i \bar{\theta} \Gamma_{(-} H_{\mu)\nu\rho} \Gamma^{\nu\rho}\sigma_3 \theta,
\end{equation}
while our transformation gives
\begin{equation}
\nonumber
\bar{\theta}\Gamma^{+}H_{\mu\nu\rho}\Gamma^{\nu\rho}\theta \rightarrow (i\bar{\eta})(i\Gamma_{-})(-H_{\mu\nu\rho}) \Gamma^{\nu\rho}(i \sigma_3) \eta = i \bar{\eta}\Gamma_{-}H_{\mu\nu\rho}\Gamma^{\nu\rho}\sigma_3 \eta
\end{equation}
matching precisely since $H$ is purely transverse.

\subsection{Fermions with dilaton and RR fields}

Having checked that the metric and B field terms behave consistently, let us investigate the fermionic terms involving the dilaton and RR fields. A priori these terms can give contributions to any part of $\hat{g}$ and $\hat{B}$. If our transformation is to work we get immediate restrictions however, due to the transformation properties of the spinor structure in $\mathcal{S}$ (see eq. \eqref {eq:Sformdefinition})
\begin{equation}
M^t \epsilon M = - b c \epsilon, \hspace{20pt} M^t \sigma_1 M = b c \sigma_1,
\end{equation}
actually leaving the type of couplings invariant. Moreover these transformations guarantee that a real $i \bar{\theta} \Gamma_M \mathcal{S} \Gamma_n \theta$ remains real, since $bc=\pm1$ and fermionic terms that are nonzero in our $\kappa$ gauge contain an even number of non-transverse $\Gamma$ matrices. This makes it impossible for the contribution of $\mathcal{S}$ to relate the double Wick rotation of $\hat{g}$, to $\hat{B}$, and vice versa.\footnote{Since $\sigma_1 \sigma_3 = - \epsilon$ and in our $\kappa$ symmetry gauge we have $\Gamma^0 \Gamma^p \theta = \theta$, we might imagine that (the $0p$ containing components of) an $F^{(n)}$ term effectively starts looking like an $F^{(n-2)}$ term with an extra $\sigma_3$. However these terms would not couple correctly to the bosons (sitting in $\hat{g}$ rather than $\hat{B}$ or vice versa), and moreover would still not pick up the required factor of $i$. Still, we could more specifically imagine a pair of forms such that $\slashed{F}^{(n)} = \pm\slashed{F}^{(n-2)}\Gamma^0 \Gamma^p$, giving $\sigma_1 \pm \epsilon \Gamma^0 \Gamma^p$ which removes terms due to our $\kappa$ gauge choice. In this particular setting the contributions to $\hat{g}$ and $\hat{B}$ take the same overall form. Nevertheless the lacking factor of $i$ remains a problem.}

From the fact that our forms cannot lead to a mixing of light cone and transverse components in $\hat{g}$ and $\hat{B}$,
\begin{equation}
\bar{\theta} \Gamma^\pm \mathcal{S} \Gamma^\mu \theta = \bar{\theta} \Gamma^\mu \mathcal{S} \Gamma^\pm \theta  = 0\,,
\end{equation}
we deduce that each nonzero term in $\mathcal{S}$ must have a single light cone index, since three or more light cone indices necessarily give zero, and two or none gives a nonzero contribution in the above. Note that these conditions also immediately imply $\hat{B}_{+-}=0$. We then have to insist these terms transform to match eqs. \eqref{eq:tfreqs2}. For the light-cone components we have
\begin{equation}
\left(\begin{array}{c}\bar{\theta} \Gamma^+ \mathcal{S} \Gamma^\pm \theta \\ \bar{\theta} \Gamma^- \mathcal{S} \Gamma^\pm \theta
\end{array}\right)
\rightarrow
\pm i \left(\begin{array}{c}-\bar{\theta} \Gamma_- \check{\mathcal{S}} \Gamma_\mp \theta\\ \bar{\theta} \Gamma_+ \check{\mathcal{S}} \Gamma_\mp \theta\end{array}\right),
\end{equation}
where
\begin{equation}
\check{\mathcal{S}} = - b c \mathcal{S}\Big|_{\begin{subarray}{l}\Gamma^{0/5} \rightarrow i\Gamma^{5/0}\\ \sigma_1 \rightarrow - \sigma_1\end{subarray}}.
\end{equation}
The purely transverse components simply give
\begin{equation}
\bar{\theta} \Gamma_\mu \mathcal{S} \Gamma_\nu \theta \rightarrow i \bar{\theta} \Gamma_\mu \check{\mathcal{S}} \Gamma_\nu \theta.
\end{equation}
In the $\hat{B}$ case we get an extra sign here due to $\sigma_3$. Taking into account the extra minus sign for the fermions in the inverse of $\hat{g}$, these transformations are compatible with eqs. \eqref{eq:tfreqs2} provided we have a set of mirror fields (packaged in $\tilde{\mathcal{S}}$) such that
\begin{equation}
\bar{\theta} \Gamma_\mathsc{m} \tilde{\mathcal{S}} \Gamma_\mathsc{n} \theta = \begin{cases} i \bar{\theta} \Gamma_\mathsc{m} \check{\mathcal{S}} \Gamma_\mathsc{n} \theta & \mathsc{m},\mathsc{n} \in \{\mu\},\\
-i\bar{\theta} \Gamma_\mathsc{m} \check{\mathcal{S}} \Gamma_\mathsc{n} \theta &  \mathsc{m},\mathsc{n} \in \{+,-\}.\end{cases}
\end{equation}
The problematic looking relative sign we can now remove by inserting $1$ in the form of $(\Gamma^0 \Gamma^p)^2$, giving
\begin{equation}
\bar{\theta} \Gamma_\mathsc{m} \tilde{\mathcal{S}} \Gamma_\mathsc{n} \theta = - i \bar{\theta} \Gamma_\mathsc{m} \Gamma^0 \Gamma^p \check{\mathcal{S}} \Gamma_\mathsc{n} \theta
\end{equation}
since $\Gamma^0 \Gamma^p$ commutes with transverse $\Gamma$ matrices but anti-commutes with light cone ones, and we have $\bar{\theta} \Gamma^0 \Gamma^p = - \bar{\theta}$. This means that everything is precisely compatible, provided the forms of the mirror background are given by
\begin{equation}
\tilde{\mathcal{S}} = -i \Gamma^0 \Gamma^p \check{\mathcal{S}} = - i b c \Gamma^0 \Gamma^p \mathcal{S}\Big|_{\begin{subarray}{l}\Gamma^{0/5} \rightarrow i\Gamma^{5/0}\\ \sigma_1 \rightarrow - \sigma_1\end{subarray}}.
\end{equation}
We can actually rewrite this very nicely. Since $\mathcal{S}$ must have one and only one light cone index, we can always write it as $\Gamma^0 N + \Gamma^p K$ for some $N$ and $K$ that do not contain $\Gamma^0$ or $\Gamma^p$. In this form it is clear however that multiplication by $\Gamma^0 \Gamma^p$ undoes the interchange of $\Gamma^0$ for $\Gamma^p$ in $\check{\mathcal{S}}$, leaving just an extra factor of $i$. In short, we have
\begin{equation}
\tilde{\mathcal{S}} = -i \Gamma^0 \Gamma^p \check{\mathcal{S}} =  b c \mathcal{S}\Big|_{\sigma_1 \rightarrow - \sigma_1}.
\end{equation}
This transformation is uniquely fixed up to the sign choice $bc=\pm1$. However, since a simultaneous sign change on all RR fields leaves the supergravity equations of motion \cite{Schwarz:1983qr} invariant, we can consider this sign choice irrelevant anyway.

We see that two backgrounds are related by a double Wick rotation provided that the metric and B field are related as in figure \ref{fig:bosmirrortf} and the dilaton and RR fields as
\begin{equation}
\label{eq:slashedRRtransformations}
\begin{aligned}
e^{\Phi} \slashed{F}^{(n)} &  \rightarrow (i)^{n-1} e^{\Phi} \slashed{F}^{(n)},
\end{aligned}
\end{equation}
where we have chosen $bc=1$. With this choice, the story unifies at the level of the underlying even degree forms (see appendix \ref{app:supergravity})
\begin{equation}
\begin{aligned}
e^{\Phi} d\slashed{C}^{(n)} &\rightarrow i^n e^{\Phi} d\slashed{C}^{(n)},\\
d\slashed{B} &\rightarrow i^2 d\slashed{B}.
\end{aligned}
\end{equation}
In short, up to a possible sign the combinations $e^\Phi \slashed{F}$ for the mirror background are identical to the originals. Of course the slashes can be removed, but then it is important to note that the relations hold between tensors with flat space indices. Our full story can now be summarized by figure \ref{fig:totalmirrortf}.
\begin{figure}
\begin{center}
\begin{tikzpicture}
\node at (-5/2,0) {{\LARGE $\mathcal{L}$}};
\node at (5/2,0) {{\LARGE $\tilde{\mathcal{L}}$}};
\node at (0,3/2) {\small $\tau \rightarrow i \tilde{\sigma},$ \hspace{3pt} $\sigma \rightarrow -i \tilde{\tau}$};
\node at (0,-3/2) {\small $g_{\rm{lc}}\rightarrow \frac{g_{\rm{lc}}}{|\det g_{\rm{lc}}|},$ \hspace{3pt} $B\rightarrow-B,$};
\node at (0,-2.2) {\small $e^{\Phi} \slashed{F}^{(n)} \rightarrow (i)^{n-1} e^{\Phi} \slashed{F}^{(n)}$};
\draw[black, thick,->, domain=-pi/3:pi/3] plot ({2*sin(\x r)},{cos(\x r)});
\draw[black, thick,<-, domain=-pi/3:pi/3] plot ({2*sin(\x r)},{-1*cos(\x r)});
\end{tikzpicture}
\end{center}
\caption{A double Wick rotation as a change of background fields. A double Wick rotation of the light cone gauge fixed world sheet theory of a Green-Schwarz string is equivalent to a change of background fields, at least to second order in the fermions. It is not clear whether these changed fields always correspond to a solution of supergravity.}
\label{fig:totalmirrortf}
\end{figure}
At this level we cannot immediately disentangle the dilaton and the RR fields.

\section{Double Wick related string backgrounds}
\label{sec:doubleWickBackgrounds}

In the previous section we found transformation laws for the background fields via an explicit double Wick rotation on the world sheet combined with an appropriate fermionic field redefinition. Before giving a few examples of pairs of backgrounds related by these transformations, let us discuss how these transformation laws can also be viewed more directly from a target space point of view.

As used in our appendix, light cone gauge fixing can be viewed as T dualizing in the $x^-$ direction, and gauge fixing the corresponding T dual field $\psi = \sigma$ in addition to $x^+ = \tau$ \cite{Kruczenski:2004cn}. It should then be possible to (formally) view a double Wick rotation on the world sheet at the target space level as a T duality in $x^-$, followed by the analytic continuation $(x^+,\psi) \rightarrow (i \tilde{\psi},-i\tilde{x}^+)$, and finally another T duality in the $\tilde{\psi}$ direction.\footnote{We would like to thank A. Tseytlin for this suggestion.} As this procedure involves analytic continuation of target space fields as well as T duality in a light cone direction however,\footnote{In principle this light cone T duality can be avoided by considering the static gauge instead of our generalized light cone gauge, leading to T dualities in the $\phi$ directions.} this procedure takes us out of the realm of real supergravity and string theory, and generically does not result in a real solution of supergravity. Furthermore, the details of this analytic continuation are subtle, as the complex `diffeomorphism' corresponding to the double Wick rotation is `improper', i.e.
\begin{equation}
\left(\begin{array}{c}\tilde{\tau}(\tilde{x}^+) \\ \tilde{\sigma} (\tilde{\psi}) \end{array}\right) = \left(\begin{array}{cc}0 & i \\-i & 0 \end{array}\right) \left(\begin{array}{c}\tau (x^+) \\ \sigma (\psi) \end{array}\right)
\end{equation}
is a transformation with determinant $-1$. From this point of view our transformation laws should be compatible with the supergravity equations of motion as follows.

Firstly we observe that precisely under our restrictions on the metric and B field, and the ones just found for the RR fields, the combination of T dualities and analytic continuation preserves reality of the background fields. Namely, T dualizing metrics and B fields that fit our restrictions cannot lead to mixing of light cone and transverse directions, which are the components that would pick up factors of $i$ under the analytic continuation. Similarly, given RR fields with a single light cone index, T duality in $x^-$ results in (type IIA) RR fields with either no $\psi$ or $+$ index or both, which hence remain real under the analytic continuation.

This sequence of T duality, analytic continuation, and T duality, should match the transformations we found based on our world sheet perspective. To avoid technical complications, we will demonstrate this explicitly for metrics diagonal in $t$ and $\phi$, and fixing the static gauge $t=\tau$, $p_\phi=1$.\footnote{This gauge is equivalent to our light cone gauge for $a=0$.} By assumption our RR fields then only have components involving $t$ or $\phi$, and upon T dualizing $\phi$ to $\psi$ we get
\begin{equation}
\left(\begin{array}{c} F_{t\ldots}\\F_{\phi \ldots} \end{array}\right) \stackrel{{\tiny \mbox{T duality}}}{\longrightarrow} \left(\begin{array}{c} F_{t\psi\ldots}\\-F_{\ldots} \end{array}\right),
\end{equation}
where the dots denote an even number of transverse indices. The analytic continuation is slightly subtle however, involving more than the simple replacement $(t,\psi) \rightarrow (i \tilde{\psi},-i\tilde{t})$. The reason for this can be seen from the bosonic type IIA supergravity action (see e.g. page 172 of \cite{Tong:2009np}). While this action is generically diffeomorphism invariant, this involves a little more work for improper diffeomorphisms such as ours, due to the Chern-Simons term $B \wedge F^{(4)}\wedge F^{(4)}$. This term behaves like a pseudoscalar under diffeomorphisms, and picks up a sign under our analytic continuation.\footnote{To see this explicitly, note that our RR fields T dualized to type IIA have either both $t$ and $\psi$ as components, or neither, where the contribution with both picks up a sign under analytic continuation.} We can fix this sign by combining our analytic continuation with a sign flip on $B$. However, to then keep the kinetic term $|\hat{F}^{(4)}|^2$ invariant, since $\hat{F}^{(4)} = F^{(4)}- C^{(1)} \wedge H$, we also need to flip the sign of either $F^{(4)}$ or $F^{(2)}$ ($C^{(1)}$). Importantly, this introduces a relative sign between $F^{(n)}$ and $F^{(n+2)}$. The result of the analytic continuation and a second T duality then becomes
\begin{equation}
\left(\begin{array}{c} F_{t\psi\ldots}\\-F_{\ldots} \end{array}\right) \stackrel{{\tiny \mbox{an. cont.}}}{\longrightarrow} \pm \left(\begin{array}{c}  F_{\tilde{t}\tilde{\psi}\ldots}\\ - F_{\ldots} \end{array}\right) \stackrel{{\tiny \mbox{T duality}}}{\longrightarrow}  \pm \left(\begin{array}{c} F_{\tilde{t}\ldots}\\F_{\tilde{\phi}\ldots}  \end{array}\right),
\end{equation}
where in the analytic continuation step we implicitly swapped the indices for a sign, and the overall $\pm$ sign refers to the degree of the form, being plus for $F^{(1)}$ and $F^{(5)}$, and minus for $F^{(3)}$, or vice versa.\footnote{Recall that we encountered the same inconsequential sign freedom in our world sheet discussion above.} Combining this with the sign flip of $B$ under the analytic continuation, and the dilaton generated by the double T duality that is clearly compatible with $e^\Phi \slashed{F} = e^{\tilde{\Phi}} \tilde{\slashed{F}}$, we see that this precisely matches the transformations we found previously from a world sheet perspective, which are therefore compatible with the supergravity equations of motion.

After this general discussion let us give some examples. While our restrictions on the metric and B field do not appear too drastic, the number of explicitly known solutions of supergravity is also not very large; we will consider $\ads$, $\rm{AdS}_3 \times \rm{S}^3 \times \rm{T}^4$ supported by a RR three form, and $\rm{AdS}_2 \times \rm{S}^2 \times \rm{T}^6$.\footnote{To have a case with a transverse B fields we could for example consider the three parameter generalization of the Lunin-Maldacena background \cite{Lunin:2005jy} constructed in \cite{Frolov:2005dj}, which fits our restrictions when only $\gamma_1$ is nonzero. Presumably the associated mirror space can be obtained by TsT transformations on mirror $\ads$; we will not pursue this interesting case here.} We distinguish mirror space quantities by tildes below.

\subsection*{$\ads$}

The metric of $\ads$ in global coordinates
\begin{equation}
ds^2 =\,-(1+\rho^2)dt^2 + \frac{d\rho^2}{1+\rho^2} + \rho^2 d\Omega_{3}+ (1-r^2) d\phi^2 + \frac{dr^2}{1-r^2} +r^2 d\Omega_{3},
\end{equation}
is precisely of the form discussed in section \ref{sec:bosmirrortf}, and its mirror companion is \cite{Arutyunov:2014cra}
\begin{equation}
\label{eq:mirrorAdS5metric}
\tilde{ds}^2 =\,-\frac{1}{1-r^2}dt^2 + \frac{d\rho^2}{1+\rho^2} + \rho^2 d\Omega_{3}+ \frac{1}{1+\rho^2}d\phi^2+ \frac{dr^2}{1-r^2} +r^2 d\Omega_{3}.
\end{equation}
Note that this metric corresponds to a direct product of two manifolds with coordinates $t$, $r$, and associated angles, and $\phi$, $\rho$, and associated angles. We have written the mirror metric with this nonstandard ordering of coordinates, so that we leave the labeling of the transverse space untouched with respect to $\ads$.

Both these spaces can be embedded in type IIB supergravity, supported by a self dual five form and dilaton. The relevant equations of motion are given in appendix \ref{app:supergravity}. For $\ads$ these equations are solved by a constant dilaton $\Phi = \Phi_0$, and a five form
\begin{equation}
\tfrac{1}{4} e^\Phi \slashed{F} = \Gamma^{01234} - \Gamma^{56789}.
\end{equation}
In line with the above discussion, mirror $\ads$ \cite{Arutyunov:2014cra} has a nontrivial dilaton
\begin{equation}\label{eq:mirrordilaton}
\tilde{\Phi}=\tilde{\Phi}_0-\frac{1}{2}\log  (1-r^2)(1+\rho^2),
\end{equation}
and its combination with the five form is precisely such that
\begin{equation}
\tfrac{1}{4}e^{\tilde{\Phi}} \tilde{\slashed{F}} = \Gamma^{01234} - \Gamma^{56789} = \tfrac{1}{4} e^\Phi \slashed{F}.
\end{equation}
Note that since the metrics differ, the physical meaning of the forms is very different between $\ads$ and its mirror version; in the mirror case the form mixes directions belonging to the two five dimensional submanifolds, while the form is just a difference of five dimensional volume forms for $\ads$.

As discussed in \cite{Arutyunov:2014cra}, mirror $\ads$ and its sigma model have some interesting properties. Firstly, the sigma model on mirror $\ads$ inherits the integrability of the (gauge fixed) $\ads$ sigma model. Also, mirror $\ads$ has a (naked) singularity at $r=1$, where the dilaton blows up correspondingly. Background singularities are not necessarily problematic for the associated sigma model however, and the integrability of the mirror sigma model and in general its link to the $\ads$ sigma model suggests nice behavior. At the same time, the unboundedness of the dilaton makes it difficult to imagine extending the relation to $\ads$ beyond the free string, which is perhaps only natural since the world sheet double Wick rotation loses its simple physical interpretation on a higher genus Riemann surface. Finally, while mirror $\ads$ admits no Killing spinors, the sigma model inherits the light cone symmetries of the $\ads$ sigma model, and hence has (at least) $\mathfrak{psu}(2|2)^{\oplus 2}$ symmetry, supersymmetry being nonlinearly realized as explained in \cite{Arutyunov:2014cra}.\footnote{The four $\mathrm{SU}(2)$s correspond to the $\mathrm{SO}(4)$ symmetries of the two three spheres in eqn. \eqref{eq:mirrorAdS5metric}.} This appears to be a new type of (string) supersymmetry without (background) supersymmetry \cite{Duff:1997qz,Duff:1998us}; it is realized within the sigma model in contrast to \cite{Duff:1997qz}, and relies on a nonlinear realization of supersymmetry, not (manifestly) on string winding modes as in \cite{Duff:1998us}.

Let us also mention that by formal T duality in $t$ and $\phi$, mirror $\ads$ becomes $\mathrm{dS}_5 \times \mathrm{H}^5$, a solution of type IIB$^*$ supergravity \cite{Hull:1998vg}. It is no accident that such T dualities give a solution of type IIB$^*$ theory, as this theory is obtained from type IIA theory by timelike T duality \cite{Hull:1998vg}. Actually, at least for cases with metrics diagonal in $t$ and $\phi$, in the static gauge it is clear that T dualities in $t$ and $\phi$ undo the two T dualities in our procedure above (one having become timelike due to the intermediate analytic continuation). With the T dualities undone, what remains of our discussion is the analytic continuation $(t,\phi) \rightarrow (i \tilde{\phi},-i\tilde{t})$. Since analytically continuing $\phi\rightarrow \pm i\tilde{t}$ turns $\mathrm{S}^5$ into $\mathrm{dS}_5$, and similarly $t \rightarrow \pm i \tilde{\phi}$ turns $\mathrm{AdS}_5$ into $\mathrm{H}^5$, this explains the appearance of $\mathrm{dS}_5 \times \mathrm{H}^5$. Note that because of the intermediate continuation, both T dualities in our procedure  involve the same type of coordinates regardless of the choice of gauge parameter $a$ (e.g. the spacelike $\phi$ and $it$ in the static gauge equivalent to $a=0$). As a result, our transformations always stay within the context of type II supergravity, though the corresponding solutions may have interesting relations to solutions of type II$^*$ supergravity.

\subsection*{$\rm{AdS}_3 \times \rm{S}^3 \times \rm{T}^4$ and $\rm{AdS}_2 \times \rm{S}^2 \times \rm{T}^6$}

Taking our $\phi$ coordinate to be the equatorial angle on the sphere, the mirror metrics associated to $\rm{AdS}_3 \times \rm{S}^3 \times \rm{T}^4$ and $\rm{AdS}_2 \times \rm{S}^2 \times \rm{T}^6$ are just the lower dimensional analogues of eq. \eqref{eq:mirrorAdS5metric} completed to ten dimensions with flat directions. We choose to label coordinates such that $\phi \sim n$, hence $p=n$, for the $\rm{AdS}_n \times \rm{S}^n$ case.

The supergravity equations of motion are solved by a constant dilaton for $\rm{AdS}_3 \times \rm{S}^3 \times \rm{T}^4$ and $\rm{AdS}_2 \times \rm{S}^2 \times \rm{T}^6$. We take $\rm{AdS}_3 \times \rm{S}^3 \times \rm{T}^4$ to be supported by the three form
\begin{equation}
\tfrac{1}{2}e^\Phi \slashed{F} = \Gamma^{012} + \Gamma^{345},
\end{equation}
while $\rm{AdS}_2 \times \rm{S}^2 \times \rm{T}^6$ is supported by the five form
\begin{equation}
\label{eq:AdS2FiveForm}
e^\Phi \slashed{F} = \Gamma^{01}\mbox{Re}(\slashed{w}_{\mathbb{C}^3}) - \Gamma^{23}\mbox{Im}(\slashed{w}_{\mathbb{C}^3}),
\end{equation}
where $w_{\mathbb{C}^3}$ is the holomorphic volume form on $T^6$ in complex coordinates. Based on our discussion above, this should then also give the solution for their mirror versions. Indeed, we can readily check that this is the case with the mirror space dilaton in both cases given by eq. \eqref{eq:mirrordilaton}.\footnote{These mirror space solutions can be independently derived by making an obvious ansatz based on the result for $\ads$. Alternately, they can also be obtained by T dualizing in $t$ and $\phi$, giving $\rm{dS}_{2/3} \times \rm{H}^{2/3} \times T^{6/4}$ which are simple solutions of type IIB$^*$ supergravity analogous to $\rm{AdS}_{2/3} \times \rm{S}^{2/3}\times \rm{T}^{6/4}$ in type IIB supergravity.}

The interesting features of mirror $\ads$ discussed above each translate directly to mirror $\rm{AdS}_3 \times \rm{S}^3 \times \rm{T}^4$ and mirror $\rm{AdS}_2 \times \rm{S}^2 \times \rm{T}^6$.

\section{Conclusions and outlook}
\label{sec:conclusions}

In this paper we have found the light cone gauge fixed action for a Green-Schwarz string to second order in the fermions, for an (almost) completely general background.\footnote{Our analysis can clearly be repeated type IIA string, replacing the doublet of Majorana-Weyl spinors by a single Majorana spinor, $\sigma_3$ by $\Gamma_{11}$, and $\mathcal{S}$ by its appropriate type IIA expression \cite{Wulff:2013kga}. Accordingly, $M$ should become a $32\times32$ matrix that generates and removes $\Gamma_{11}$, and may in fact nicely combine with the automorphism that exchanges $\Gamma^{0/5}$ for $i \Gamma^{5/0}$; we do not expect other surprises. The resulting transformations will of course need to be compatible with our type IIB transformations under T duality.} We used this action to derive a set of transformation laws for the dilaton and RR fields that combined with the set of already known transformation laws for the metric and B field are equivalent to doing a double Wick rotation on the world sheet of a light cone gauge fixed string. We discussed how these transformations are compatible with the supergravity equations of motion.
With our general results we explicitly proved that the mirror models associated to $\ads$ and its lower dimensional analogues are in fact strings on the associated mirror spaces, at least to second order in fermions. These mirror spaces have interesting features, inheriting integrability as well as some supersymmetry at the level of the sigma model through the double Wick rotation, yet being singular and nonsupersymmetric solutions of supergravity.

\subsection*{More general spaces}

We would like to apply our ideas to other spaces as well, and perhaps generate interesting new string backgrounds. Many spaces that come to mind do not fit the form of our metric or B field however. For instance, the standard way to fix a light cone gauge on $\rm{AdS}_4 \times \mathbb{CP}^3$ involves an angle on $\mathbb{CP}^3$ which results in a metric that is not block diagonal as in eq. \eqref{eq:generalmetric} (see e.g. \cite{Klose:2010ki}). The same is true for $\rm{AdS}_3 \times \rm{S}^3 \times \rm{S}^3 \times \rm{S}^1$ (see e.g. \cite{Babichenko:2009dk}), while for $\rm{AdS}_3 \times \rm{S}^3 \times \rm{M}^4$ supported by a mixture of RR and NSNS three forms the B field has components mixing light cone and transverse directions, or would break the light cone isometries (see e.g. \cite{Sundin:2014ema}). The Lunin-Maldacena background \cite{Lunin:2005jy} has a similarly obstructive B field. So let us relax our restrictions and see what happens.

From the general result of appendix \ref{app:lightconegaugefixing} we can read off the light cone gauge fixed bosonic action on any space with light cone isometries and $B_{+-}=0$. As before, we would like to do a double Wick rotation in this action and attempt to interpret the result as an action of the same form. This action naturally splits into two pieces
\begin{equation}
S = - T \int \sqrt{-G} + \frac{1}{2} E\, ,
\end{equation}
and our approach applies to each piece independently. The $E$ piece is simple but representative and in the present case reduces from eq. \eqref{eq:generalE} to
\begin{equation}
E =  2\frac{g^{+-}}{g^{++}} - \frac{2}{g_{--}}\left(\mathring{B}_{\mu\nu} \dot{x}^\mu x^{\prime \nu} + g_{-\mu} \dot{x}^\mu - \mathring{B}_{+\mu} x^{\prime \mu}\right),
\end{equation}
where $\mathring{B}_{+\mu} = g_{--} B_{+\mu} - g_{+-}B_{-\mu}$ and $\mathring{B}_{\mu\nu} = g_{--} B_{\mu\nu} - g_{-\mu}B_{-\nu} + g_{-\nu}B_{-\mu}$. Clearly a double Wick rotation introduces factors of $i$ here, and in order to reinterpret this as another $E$ term we would need to consider $g$ and $B$ with imaginary components.\footnote{We could have read this off at the linearized level using the action of section \ref{sec:fermmirortf}. In fact, this situation is entirely similar to the discussion of section \ref{sec:fermmirortf} where purely transverse RR forms would generate nonzero $\hat{g}_{\pm \mu}$ or $\hat{B}_{\pm \mu}$.} $\sqrt{-G}$ shows similar behaviour, and all of this matches the T duality discussion of the previous section. In other words, staying in the realm of real metrics and B fields we cannot straightforwardly extend our considerations. However, as the mirror model for e.g. $\rm{AdS}_4 \times \mathbb{CP}^3$ has a unitary S-matrix, there may exists a (nonlocal) field redefinition that removes the problematic looking imaginary terms. What form it should precisely take, and whether the result can be interpreted as a string sigma model is an interesting open problem. Attempting to extend the deformation of \cite{Delduc:2013qra} to the coset model for $\rm{AdS}_4 \times \mathbb{CP}^3$ \cite{Arutyunov:2008if,Stefanski:2008ik} may shed light on this question.\footnote{(Models based on) even dimensional bosonic spaces presumably behave qualitatively different under this deformation however, so perhaps we should not expect to find mirror duality between real string backgrounds here.}

\subsection*{More general mirror spaces}

Another way we could attempt to generalize our considerations is to attempt to fix a light cone gauge in a different way. Let us explain this by means of example. In the previous section we consider the mirror versions of $\rm{AdS}_{2/3} \times \rm{S}^{2/3} \times \rm{T}^{6/4}$, taking the conventional equatorial angle $\phi$ on $\rm{S}^{2/3}$ to be part of our light cone coordinates. While this is the natural choice,\footnote{This choice preserves the most manifest supersymmetry, allows for an exact S-matrix approach (see e.g. \cite{Borsato:2014exa,Hoare:2014kma}, or the review \cite{Sfondrini:2014via}), and hence gives the appropriate mirror theory for the thermodynamic Bethe ansatz.} we are allowed to fix a light cone gauge using e.g. an angle $\varphi$ on $\rm{T}^{6/4}$ without violating our restrictions on the metric if we so desire. However, this changed interpretation of coordinates means that the three form now violates our restriction that it should not contribute to $\hat{g}_{\pm \mu}$ and $\hat{B}_{\pm \mu}$. Similarly to the discussion above, also here there might be a field redefinition that removes the resulting imaginary looking terms. In any case, the form also results in a nonzero $\hat{B}_{+-}$ which on its own already takes us out of the realm of our present paper. It would be very interesting to have a background with two different sets of light cone coordinates fitting our full set of restrictions.

\subsection*{Strings on deformed $\rm{AdS}_{n} \times \rm{S}^{n} \times \rm{T}^{10-2n}$}

As mentioned before, our work was inspired by the mirror duality \cite{Arutynov:2014ota} of the deformed $\ads$ sigma model of \cite{Delduc:2013qra}. It would be great to use the results of this paper to explicitly prove this mirror duality. However, it is still not clear whether the metric and B field of the deformed sigma model \cite{Arutyunov:2013ega} can be embedded in supergravity, and whether such an embedding would correspond precisely to the deformed sigma model. In lower dimensions there has been progress in this direction, which raises interesting questions.

A supergravity solution for deformed $\rm{AdS}_{2} \times \rm{S}^{2}$ and its uplift in ten dimensions for $\left(\rm{AdS}_{2} \times \rm{S}^{2}\right)_{\eta(\kappa)} \times \rm{T}^{6}$ have been found in \cite{Lunin:2014tsa}. This solution contains a free parameter $a$ and the conjecture of \cite{Lunin:2014tsa} is that it matches the coset model for a particular choice of $a(\kappa)$. Attempts to combine our results with the background of \cite{Lunin:2014tsa} raises some questions. Based on the deformed $\ads$ coset model we would expect to find mirror duality for $\left(\rm{AdS}_{2} \times \rm{S}^{2}\right)_{\eta(\kappa)} \times \rm{T}^{6}$. With a full background we can attempt to check this explicitly. At the bosonic level mirror duality works immediately following \cite{Arutyunov:2014cra}, but at the fermionic level we find a three form that does not fit our restrictions. This means that a double Wick rotation would not result in a manifestly real action. Related to the discussion above, it may be possible to resolve this by some subtle field redefinition however. Other, less satisfying possibilities are of course that for some unforseen reason, though present in the bosonic model, mirror duality is not present in the full deformed $\rm{AdS}_2$ sigma model, or that the sigma model is simply not described by this supergravity solution. Of course, it could also be the extension of the deformed coset model for $\rm{AdS}_{2} \times \rm{S}^{2}$ to a ten dimensional background that generates friction on either of these two points. Consolidating these facts, and thereby hopefully proving mirror duality, is an interesting problem. The supergravity solution of \cite{Lunin:2014tsa} does appear to reduce to our mirror solution of eqs. \eqref{eq:mirrordilaton} and \eqref{eq:AdS2FiveForm} in an appropriately scaled $\kappa\rightarrow\infty$ limit.\footnote{Note that to match our mirror solution with the $\kappa\rightarrow\infty$ limit of the result of \cite{Lunin:2014tsa} we need to take $c_1=-c_2=1/\sqrt{2}$ there. Our undeformed $\rm{AdS}_{2} \times \rm{S}^{2}\times \rm{T}^{6}$ matches the solution at $\kappa=0$ with $c_1=c_2=1/\sqrt{2}$ instead.}

\section*{Acknowledgements}

The authors would like to thank R. Borsato, S. Frolov, B. Hoare, M. Magro, and A. Tseytlin for discussions, and R. Borsato, S. Frolov, and especially A. Tseytlin for comments on the paper. S.T. is supported by the Einstein Foundation Berlin in the framework of the research project "Gravitation and High Energy Physics" and acknowledges further support from the People Programme (Marie Curie Actions) of the European Union's Seventh Framework Programme FP7/2007-2013/ under REA Grant Agreement No 317089 (GATIS). The work of G.A. is supported by the German Science Foundation (DFG) under the Collaborative Research Center (SFB) 676 Particles, Strings and the Early Universe.

\appendix

\section{Appendices}

\subsection{$\Gamma$ matrices, $\kappa$ symmetry, and other conventions}
\label{app:Gammamatricesandkappafixing}

Target space (curved) indices are denoted by upper case Latin letters, frequently split over light cone indices $+$ and $-$, and transverse indices denoted by lower case Greek letters such as $\mu$ and $\nu$ nearer the end of the alphabet. World sheet indices are denoted by lower case Greek letters from the front of the alphabet, such as $\alpha$ and $\beta$. We use $\gamma^{\alpha\beta} = \sqrt{-h} h^{\alpha\beta}$ to denote the Weyl-invariant combination of the world sheet metric $h^{\alpha\beta}$.  We work in `mostly plus' conventions where $\eta = \mbox{diag}(-1,1,\ldots1)$, and we label coordinates in flat space by the numbers $0,\ldots,9$, generically denoted by lower case Latin letters. We define the flat epsilon tensor in both two and ten dimensions with canonically ordered upper indices to be one, i.e. $\epsilon^{01\ldots}=1$.

Given a metric of the form of eq. \eqref{eq:generalmetric}, we can take a block diagonal vielbein, meaning we can and will talk about transverse flat directions as well; $0$ and $p$ denote the flat directions associated to $t$ and $\phi$ respectively, the rest is transverse.

We can work in a Majorana basis where all $\Gamma$-matrices are purely real,
\begin{equation}
\label{eq:gammamatrixtransposition}
(\Gamma^0)^t = - \Gamma^0\,, \hspace{25pt} (\Gamma^i)^t = \Gamma^i \, \, \mbox{ for }\, i = 1,\ldots,9,
\end{equation}
and
\begin{equation}
\Gamma_{11} = \left(\begin{array}{cc} I_{16} & 0 \\ 0 & -I_{16} \end{array}\right).
\end{equation}
A Majorana-Weyl spinor of positive chirality satisfies $\theta^t C = \bar{\theta} \equiv \theta^\dagger \Gamma^0$ and $\tfrac{1}{2}(1+\Gamma_{11}) \theta = \theta$, where $C$ is the charge conjugation matrix we can take to be $\Gamma^0$, and is thus given by
\begin{equation}
\theta = \left(\begin{array}{c} \psi \\ 0 \end{array} \right),
\end{equation}
where $\psi$ is 16 dimensional and real.\footnote{We can make contact with the 16 dimensional conventions of \cite{Wulff:2013kga} by taking
\begin{align}
\Gamma^0 & = i \sigma_2 \otimes 1,\\
\Gamma^a & = \sigma_1 \otimes \gamma^a,
\end{align}
where $\gamma^b$, $b=1,\ldots,8$ denotes a set of eight dimensional euclidean $\gamma$ matrices, and $\gamma^9$ is equal to their product. In this basis the action of the $\Gamma$ matrices on a positive chirality Majorana-Weyl spinor reduces to that of the $\gamma$ matrices with an effective $\gamma^0$ of $1$. Note that $\gamma^{a\ldots k}$ is not simply the product of $\gamma$ matrices; its sign is fixed by comparing to $\Gamma^{a \ldots k}$.}

To fix $\kappa$ symmetry we introduce the matrices
\begin{equation}
G^\pm = \frac{1}{\sqrt{2}} (\Gamma^0 \pm \Gamma^p),
\end{equation}
which satisfy
\begin{align}
G^\pm G^\pm & \, =0,
\end{align}
and
\begin{equation}
\label{eq:gammaplusminussum}
-\frac{1}{2} (G^+G^- + G^-G^+)  = 1.
\end{equation}
These matrices are not the target space light cone gamma matrices $\Gamma^\pm$. Both $G$ matrices have rank sixteen, as they are nilpotent but sum to $\Gamma^0$ which is nonsingular. Note that
\begin{equation}
(G^\pm)^t = - G^\mp.
\end{equation}
We fix $\kappa$ symmetry by imposing
\begin{equation}
G^+ \theta = 0.
\end{equation}
With a metric of the form of eq. \eqref{eq:generalmetric} we see that
\begin{equation}
\bar{\theta} \Gamma_\mu \ldots \Gamma_\nu \theta \equiv
\theta^t \Gamma^0 \Gamma_\mu \ldots \Gamma_\nu \theta = 0,
\end{equation}
for any number of transverse $\Gamma$ matrices (including zero), by inserting $1$ from eq. \eqref{eq:gammaplusminussum} and either having $G^+$ acting on $\theta$ to the right, or $G^-$ ($G^+$) on $\theta^t$ ($\bar{\theta}$) to the left. We similarly get zero with two light cone $\Gamma$ matrices and an arbitrary number of transverse ones. We also have
\begin{equation}
\label{eq:thetabarviagamma5}
\theta^t \gamma^- = 0 \, \Rightarrow \,\bar{\theta} = \theta^t \Gamma^0 = \theta^t \Gamma^p.
\end{equation}
For completeness we note
\begin{align}
\Gamma^0 \Gamma^p \theta = \theta,
\end{align}
and
\begin{equation}
\bar{\theta} \Gamma^0 \Gamma^p = - \bar{\theta}.
\end{equation}

\subsection{Light cone gauge fixing}
\label{app:lightconegaugefixing}

In this appendix we will consider almost completely generic spaces for which we can fix a light cone gauge, meaning that the components depend on the transverse coordinates $x^\mu$ only
\begin{equation}
\label{eq:metricgen}
\mathrm{ds}^2 = g_{\mathsc{mn}}(x^\mu) \,  d x^\mathsc{m} d x^\mathsc{n}
\end{equation}
and, including a B field, and fermions to second order, actions like
\begin{align}
S =-\tfrac{T}{2} \int d \sigma d \tau (& \gamma^{\alpha \beta} \hat{g}_{\mathsc{mn}} \partial_\alpha  x^\mathsc{m} \partial_\beta  x^\mathsc{n} - \epsilon^{\alpha\beta} \hat{B}_{\mathsc{mn}} \partial_\alpha x^\mathsc{m}  \partial_\beta  x^\mathsc{n} \\
&+ i\gamma^{\alpha \beta} \partial_\alpha x^\mathsc{m} \bar{\theta} \Gamma_\mathsc{m} \partial_\beta\theta - i\epsilon^{\alpha \beta} \partial_\alpha x^\mathsc{m} \bar{\theta} \Gamma_\mathsc{m} \sigma_3 \partial_\beta \theta )\,
\end{align}
The only restriction we impose is that $B_{t\phi}=B_{+-}=0$. In the rest of this appendix we will avoid clutter by dropping the hats on $\hat{g}$ and $\hat{B}$, but of course their fermionic contributions should still be carried along.

To fix a light cone gauge we will follow the approach of \cite{Kruczenski:2004cn}, nicely summarized in \cite{Zarembo:2009au}. Our results will generalize those of \cite{Zarembo:2009au,Klose:2006zd,Arutyunov:2009ga} to almost completely generic bosonic backgrounds, and moreover include fermions to second order. We begin by gauging the isometry
\begin{equation}
x^- \rightarrow x^- + \xi,
\end{equation}
by adding a gauge field $A$ to the Lagrangian, and imposing that it is flat so that we actually do nothing. That is, we make the minimal replacement $d x^- \rightarrow dx^- + A$ and add a term enforcing the flatness of $A$, giving
\begin{align}
\label{eq:gaugedaction}
(-T)^{-1} (\mathcal{L}_A- \left.\mathcal{L}_A \right|_{A=0} ) = &\, \frac{1}{2} g_{--}  A_\alpha A^{\alpha}+ A_\alpha \gamma^{\alpha\beta}( g_{-\mathsc{n}} \partial_\beta x^\mathsc{n} + \tfrac{i}{2} \bar{\theta} \Gamma_- \partial_\alpha \theta)  \\
& - A_\alpha \epsilon^{\alpha \beta}(B_{-\mathsc{n}} \partial_\beta x^\mathsc{n} + \tfrac{i}{2}\bar{\theta} \Gamma_{-} \sigma_3 \partial_\beta \theta) + \psi \epsilon^{\alpha\beta}\partial_\alpha A_\beta \nonumber
\end{align}
where $\psi$ is the lagrange multiplier for the flatness of $A$. We fixed the sign of $\psi$ such that the equation of motion for $A_\tau$ gives us $\psi^\prime = p_-/T$. We now integrate out the $A$ field to get
\begin{equation}
\mathcal{L} = \left. \mathcal{L}_A \right|_{A=0} + \frac{T}{2g_{--}} \left(\gamma^{\alpha\beta} (g_{-\mathsc{n}} \partial_\alpha x^\mathsc{n} + \tfrac{i}{2}\bar{\theta} \Gamma_- \partial_\alpha \theta)-\epsilon^{\alpha \beta}(B_{-\mathsc{n}} \partial_\beta x^\mathsc{n} + \tfrac{i}{2} \bar{\theta} \Gamma_{-} \sigma_3 \partial_\beta \theta - \partial_\beta \psi)\right)^2\,, \nonumber
\end{equation}
where we integrated by parts on the $\psi$ term. The square is of course meant to be appropriately contracted using $\gamma$. Now we notice that any term in the original action with one or more minus indices cancels precisely against one in the square, as it has to be since we gauged a shift symmetry in the $x^-$ direction. We then get \footnote{Recall we assume $B_{+-}=0$.}
\begin{align}
\mathcal{L} = &\left. \mathcal{L}_A \right|_{A=0} + \frac{T}{2g_{--}} \left(\gamma^{\alpha\beta} (g_{-\mathsc{a}} \partial_\alpha x^\mathsc{a} + \tfrac{i}{2}\bar{\theta} \Gamma_- \partial_\alpha \theta)-\epsilon^{\alpha \beta}(B_{-\mu} \partial_\beta x^\mu + \tfrac{i}{2} \bar{\theta} \Gamma_{-} \sigma_3 \partial_\beta \theta - \partial_\beta \psi)\right)^2 \nonumber
\end{align}
where the indices $\mathsc{a}$ and $\mathsc{b}$ run over $\{+,\mu\}$ but not $-$, also implicitly in $\left. \mathcal{L}_A \right|_{A=0}$, and we have dropped $T \epsilon^{\alpha \beta} \partial_\alpha x^- \partial_\beta \psi$ which is zero upon integration by parts. To write it out in components let us introduce the `T dual' quantities
\begin{align}
\mathring{g}_{\mathsc{ab}} =&\, g_{--} g_{\mathsc{ab}} - g_{-\mathsc{a} } g_{-\mathsc{b}}\,,\\
\mathring{\Gamma}_\mathsc{a} = & \, g_{--} \Gamma_\mathsc{a} - g_{-\mathsc{a}} \Gamma_{-}
\end{align}
to write the Lagrangian as
\begin{align}
- \frac{2}{T}\mathcal{L} = &\,\frac{\gamma^{\alpha \beta}}{g_{--}} \Bigg(\mathring{g}_{++} \partial_\alpha x^+ \partial_\beta x^+ + 2 \mathring{g}_{+\mu} \partial_\alpha x^+ \partial_\beta x^\mu + \mathring{g}_{\mu\nu} \partial_\alpha x^\mu \partial_\beta x^\nu \nonumber \\
& \, \hspace{30pt} + i  \partial_{\alpha} x^{+} \bar{\theta} \mathring{\Gamma}_+ \partial_\beta \theta + i\partial_{\alpha} x^{\mu} \bar{\theta} \mathring{\Gamma}_\mu \partial_\beta \theta \\
& \, \hspace{30pt} + (B_{-\mu} \partial_\alpha  x^\mu + \tfrac{i}{2} \bar{\theta} \Gamma_{-} \sigma_3 \partial_\alpha \theta -\partial_\alpha \psi) (B_{-\nu} \partial_\beta x^\nu + \tfrac{i}{2} \bar{\theta} \Gamma_{-} \sigma_3 \partial_\beta \theta -\partial_\beta \psi) \Bigg)\nonumber\\
& \, + \frac{2 \epsilon^{\alpha \beta}}{g_{--}} (g_{-\mathsc{a}} \partial_\alpha x^\mathsc{a} + \tfrac{i}{2}\bar{\theta} \Gamma_- \partial_\alpha \theta)(B_{-\mu} \partial_\beta x^\mu + \tfrac{i}{2} \bar{\theta} \Gamma_{-} \sigma_3 \partial_\beta \theta - \partial_\beta \psi) \nonumber \\
& \,- \epsilon^{\alpha \beta} (2B_{+\mu} \partial_\alpha x^+ \partial_\beta x^\mu + B_{\mu\nu} \partial_\alpha x^\mu \partial_\beta x^\nu + i \partial_\alpha x^+ \bar{\theta} \Gamma_+ \sigma_3 \partial_\beta \theta + i \partial_\alpha x^\mu \bar{\theta} \Gamma_{\mu} \sigma_3 \partial_\beta \theta),\nonumber
\end{align}
At this point we want to go to the Nambu-Goto form of the Lagrangian. Writing the action as (recall $\gamma^{\alpha \beta} = \sqrt{-h} h^{\alpha \beta}$)
\begin{equation}
S = -\frac{T}{2} \int \sqrt{-h} h^{\alpha \beta} G_{\alpha \beta} + E\,,
\end{equation}
the equation of motion for the world sheet metric then reads
\begin{equation}
- T \sqrt{-h} (G_{\alpha \beta} - \frac{1}{2} h_{\alpha \beta} h^{\gamma \delta} G_{\gamma \delta}) = 0\,,
\end{equation}
giving
\begin{equation}
G = \mbox{det}(G_{\alpha \beta}) = \frac{h}{4}  (h^{\gamma \delta} G_{\gamma \delta})^2\,.
\end{equation}
Solving for $\sqrt{-h}$ and substituting back in the action gives
\begin{equation}
\label{eq:actionintermsofGandE}
S = - T \int \sqrt{-G} + \frac{1}{2} E\, .
\end{equation}
Imposing the gauge choice\footnote{We want to fix a uniform light cone gauge with $p_- = 1$ in units the string tension, and as mentioned before the equation of motion for $A$ gives $\psi^\prime = p_-/T$. Put differently, we fix our conventions to agree with the bosonic gauge fixed Lagrangian found in \cite{Arutynov:2014ota}, which involved a rescaling of $\sigma$ by $T$.}
\begin{equation}
x^+ = \tau, \hspace{20pt} \psi = \sigma \,,
\end{equation}
we find
\begin{align}
g_{--}^2 G = &\,\Bigg(\mathring{g}_{++}  + 2 \mathring{g}_{+\mu} \dot{x}^\mu + \mathring{g}_{\mu\nu} \dot{x}^\mu \dot{x}^\nu
+i  \bar{\theta} \mathring{\Gamma}_+ \dot{\theta} + i\dot{x}^{\mu} \bar{\theta} \mathring{\Gamma}_\mu \dot{\theta}+ (B_{-\mu} \dot{x}^\mu + \tfrac{i}{2} \bar{\theta} \Gamma_{-} \sigma_3 \dot{\theta})^2\Bigg)\nonumber \\
\label{eq:generalG}
&\, \times \Bigg(\mathring{g}_{\mu\nu}x^{\prime\mu} x^{\prime\nu}  + i x^{\prime \mu} \bar{\theta} \mathring{\Gamma}_\mu \theta^\prime + (B_{-\mu} x^{\prime\mu} + \tfrac{i}{2} \bar{\theta} \Gamma_{-} \sigma_3\theta^\prime-1)^2 \Bigg)\\
& \, -\Bigg(\mathring{g}_{+\mu}x^{\prime\mu} + \mathring{g}_{\mu\nu} \dot{x}^\mu x^{\prime \nu} +  \frac{i}{2} \bar{\theta} \mathring{\Gamma}_+ \theta^\prime + \frac{i}{2} (\dot{x}^{\mu} \bar{\theta} \mathring{\Gamma}_\mu \theta^\prime + x^{\prime\mu} \bar{\theta} \mathring{\Gamma}_\mu \dot{\theta}) \nonumber\\
& \, \hspace{30pt} + (B_{-\mu} \dot{x}^\mu + \tfrac{i}{2} \bar{\theta} \Gamma_{-} \sigma_3 \dot{\theta}) (B_{-\nu} x^{\prime\nu} + \tfrac{i}{2} \bar{\theta} \Gamma_{-} \sigma_3 \theta^\prime-1) \Bigg)^2, \nonumber
\end{align}
and
\begin{align}
\label{eq:generalE}
E = & \,\frac{2}{g_{--}} (g_{+-} + g_{-\mu} \dot{x}^\mu + \tfrac{i}{2}\bar{\theta} \Gamma_- \dot{\theta})(B_{-\mu} x^{\prime \mu} + \tfrac{i}{2} \bar{\theta} \Gamma_{-} \sigma_3 \theta^\prime - 1) \\
& \,  - (2B_{+\mu} x^{\prime\mu}+ 2B_{\mu\nu} \dot{x}^\mu x^{\prime\nu} + i \bar{\theta} \Gamma_{+} \sigma_3 \theta^\prime+ i \dot{x}^\mu \bar{\theta} \Gamma_{\mu} \sigma_3 \theta^\prime) \nonumber \\
& \, - \frac{2}{g_{--}} (g_{-\mu} x^{\prime \mu} + \tfrac{i}{2}\bar{\theta} \Gamma_- \theta^\prime)(B_{-\mu} \dot{x}^\mu + \tfrac{i}{2} \bar{\theta} \Gamma_{-} \sigma_3 \dot{\theta}) + i x^{\prime\mu} \bar{\theta} \Gamma_{\mu} \sigma_3 \dot{\theta}\nonumber\\
= & \, -\frac{2}{g_{--}} (g_{+-}  + \mathring{B}_{+\mu} x^{\prime \mu} + g_{-\mu} \dot{x}^\mu + \mathring{B}_{\mu\nu} \dot{x}^\mu x^{\prime \nu})\nonumber\\
& \, + \frac{1}{g_{--}}(i B_{-\mu} \bar{\theta} \Gamma_- (\dot{\theta} x^{\prime \mu} - \theta^\prime \dot{x}^{\mu})+ ig_{-\mu} \bar{\theta} \Gamma_{-} \sigma_3 (\theta^\prime \dot{x}^\mu-\dot{\theta} x^{\prime \mu})) \nonumber\\
& \,  - \frac{1}{g_{--}}( i\bar{\theta} \Gamma_- \dot{\theta}  - i g_{+-}\bar{\theta} \Gamma_{-} \sigma_3 \theta^\prime) - i \dot{x}^\mu \bar{\theta} \Gamma_{\mu} \sigma_3 \theta^\prime + i x^{\prime\mu} \bar{\theta} \Gamma_{\mu} \sigma_3 \dot{\theta}- i \bar{\theta} \Gamma_{+} \sigma_3 \theta^\prime, \nonumber
\end{align}
where $\mathring{B}_{+\mu} = g_{--} B_{+\mu} - g_{+-}B_{-\mu}$ and $\mathring{B}_{\mu\nu} = g_{--} B_{\mu\nu} - g_{-\mu}B_{-\nu} + g_{-\nu}B_{-\mu}$, and we drop terms of higher order in fermions. Putting these expressions in the action \eqref{eq:actionintermsofGandE}, including the fermions absorbed in $g$ and $B$, and expanding the fermions under the square root directly gives the light cone action of a string on (almost) any background to second order in fermions; we will not present the resulting lengthy expressions explicitly.

\subsection*{Simpler backgrounds}

Restricting ourselves to backgrounds of the form of eq. \eqref{eq:generalmetric} whereby $g_{\pm \mu}$ and $B_{\pm \mu}$ automatically become of second order in fermions, imposing our $\kappa$ gauge choice, and dropping terms higher order in fermions, the expression simplify considerably. To lowest nontrivial order the T dual quantities simplify to
\begin{align*}
\mathring{g}_{\mu\nu} =&\, g_{--} g_{\mu\nu}, & \mathring{B}_{+\mu} =& \, g_{--} B^+_{\,\,\,\mu}/g^{++}, \\
\mathring{g}_{+\mu} =&\, -g_{--} g^{+\nu} g_{\nu\mu}/g^{++}, & \mathring{B}_{\mu\nu} =&  g_{--} B_{\mu\nu},\\
\mathring{g}_{++} =&\, g_{--}/g^{++},& \mathring{\Gamma}_+ = & \, g_{--} \Gamma^+/g^{++},\\
& & \mathring{\Gamma}_\mu = & \, g_{--} \Gamma_\mu,
\end{align*}
since $g^{+-}/g^{++} = - g_{+-}/g_{--}$ (appropriately expanded in fermions). $g^{+\mu}$ gets a minus sign as we would like to preserve notation where $g$ with upper indices is the inverse of $g$ with lower indices, but in this case the component is purely second order in fermions. Indices are now raised and lowered with the regular metric of course. Using this we can simplify $E$ to get
\begin{align}
E = 2\frac{g^{+-}}{g^{++}} -2 B_{\mu\nu} \dot{x}^\mu x^{\prime \nu} -2\frac{B^{+\mu}}{g^{++}} x^{\prime}_{\,\,\mu} - 2 \frac{g_{-\mu}}{g_{--}} \dot{x}^\mu  - \frac{i}{g_{--}}\bar{\theta} \Gamma_- \dot{\theta} -  \frac{i}{g^{++}} \bar{\theta} \Gamma^{+} \sigma_3 \theta^\prime,
\end{align}
while $G$ becomes
\begin{align}
G = &\,\frac{1}{g_{--}g^{++}}\Bigg[\Bigg(1 + g^{++} \dot{x}_\mu \dot{x}^\nu
 -2 g^{+\mu} \dot{x}_\mu +i  \bar{\theta} \Gamma^+ \dot{\theta}\Bigg)\Bigg(1 + g_{--}x^{\prime}_{\,\mu} x^{\prime\mu} - 2B_{-\mu} x^{\prime\mu} - i \bar{\theta} \Gamma_{-} \sigma_3\theta^\prime \Bigg)\nonumber\\
& \, \hspace{10pt}-\Bigg( g^{++} \dot{x}_\mu x^{\prime \mu} - 2g^{+\mu}x^{\prime}_{\mu} +  i\bar{\theta} \Gamma^+ \theta^\prime  \Bigg)\Bigg(g_{--} \dot{x}_\mu x^{\prime \mu}  - 2B_{-\mu} \dot{x}^\mu - i\bar{\theta} \Gamma_{-} \sigma_3 \dot{\theta}\Bigg)\Bigg].
\end{align}
Expanding $\sqrt{-G}$ to second order in fermions we get
\begin{align}
\sqrt{-G} = \sqrt{Y^{(0)}}+\frac{1}{2\sqrt{Y^{(0)}}}&\frac{1}{g_{--}g^{++}}\Big[(2B_{-\mu} x^{\prime \mu} + i \bar{\theta} \Gamma_- \sigma_3 \theta^\prime)\Anot + (2 g^{+\mu} \dot{x}_{\mu} - i \bar{\theta} \Gamma^+ \dot{\theta}) A \\
 \nonumber & - \left(g_{--}(2g^{+\mu} x^{\prime}_{\mu} - i \bar{\theta} \Gamma^+ \theta^\prime) + g^{++}(2B_{-\mu} \dot{x}^\mu + i \bar{\theta} \Gamma_- \sigma_3 \dot{\theta})\right)\dot{x}_\mu x^{\prime\mu}\Big],
\end{align}
where $A$, $C$, and $Y^{(0)}$ are given in eqs. (\ref{eq:Y0def},\ref{eq:Adefinition}), and $B_{-\mu}$ and $g^{+\mu}$ are second order in fermions and have hats in the main text. The fermionic pieces of $\sqrt{-G}$ and $E$ give eqs. (\ref{eq:Lfa},\ref{eq:Lfb}). We have checked these expressions by explicit light cone gauge fixing in the Hamiltonian formalism as well.

\subsection{The spin connection}

\label{app:spinconnection}

Here we discuss the spin connection for metrics of the form of eq. \eqref{eq:generalmetric}. In general we have
\begin{equation}
\omega^{ab}_\nu = e^a_\alpha (\partial_\nu e^{\alpha b} + e^{\sigma b} \Gamma^{\alpha}_{\sigma\nu})\,,
\end{equation}
where $\Gamma$ is the Christoffel connection
\begin{equation}
\Gamma^\alpha_{\sigma \nu} = \frac{1}{2} g^{\alpha\beta}(\partial_\sigma g_{\beta\nu} + \partial_\nu g_{\beta\sigma} - \partial_\beta g_{\sigma\nu})\,.
\end{equation}
The Christoffel connection has a simpler expression in case any of its components involve $+$ or $-$
\begin{align}
\Gamma^{\chi}_{\psi \nu} = &\,\frac{1}{2} g^{\chi\zeta} \partial_\nu g_{\zeta \psi}\,,\\
\Gamma^{\mu}_{\chi \psi} = &\,-\frac{1}{2} g^{\mu\nu} \partial_\nu g_{\chi \psi}\,,
\end{align}
where $\chi$, $\psi$ and $\zeta$ run over $+$ and $-$, and other indices are transverse. For other indices we will not need a more specific form. The spin connection also has a simple form when its index is $\chi$
\begin{equation}
\label{eq:spinconnectionchi}
\omega_\chi^{ab} = -e^{\nu [a|} e^{\zeta |b]}\partial_\nu g_{\zeta\chi}.
\end{equation}
When its index is transverse, a simple computation shows that the spin connection has only nonzero components in the transverse space; it is formally equal to the spin connection on the eight dimensional transverse space.

By similar steps as those used to arrive at the transformations \eqref{eq:gammapmtf}, from eq. \eqref{eq:spinconnectionchi} we see that
\begin{equation}
\left(\begin{array}{c}\slashed{\omega}^\pm\\ \slashed{\omega}_\pm
\end{array}\right)\rightarrow \mp  i \left(\begin{array}{c}\slashed{\omega}_{\mp}\\
-\slashed{\omega}^{\mp}
\end{array}\right).
\end{equation}
under the change \eqref{eq:automorphism}, where the right hand side is rewritten in terms of the mirror vielbein.

\subsection{Type IIB supergravity}

\label{app:supergravity}

The field content of type IIB supergravity is given by a set of NSNS and RR fields. The NSNS fields consist of the metric $g$, the dilaton $\Phi$, and the antisymmetric two form $B$, while the RR fields consist of the axion $C^{(0)}$, the antisymmetric two form $C^{(2)}$, and the antisymmetric four form $C^{(4)}$. These forms are related to the field strengths $F$ as
\begin{equation}
\begin{aligned}
F^{(1)}&=dC^{(0)} , \\
F^{(3)}&=dC^{(2)} -C^{(0)} dB , \\
F^{(5)}&=dC^{(4)}+5(B \wedge d C^{(2)}- C^{(2)} \wedge dB) ,
\end{aligned}
\end{equation}
while the NSNS three form field strength $H=dB$. We will drop the explicit labels $(n)$ on $C^{(n)}$ and $F^{(n)}$ when there is no chance of confusion.

The concrete backgrounds we consider in this paper have no B field or axion. The (reduced) supergravity equations of motion are then given by
\begin{equation}
4 \nabla^2 \Phi - 4(\nabla \Phi)^2 = R,
\end{equation}
for the dilaton,
\begin{equation}
\partial_\mathsc{m} \left(\sqrt{-g} F^{\mathsc{mpq}(\mathsc{rs})}\right) =0,
\end{equation}
for the three and five form, and
\begin{equation}
R_{\mathsc{m}\mathsc{n}} = - 2 \nabla_\mathsc{m} \nabla_\mathsc{n} \Phi
+\frac{1}{4} e^{2\Phi}\left(F_\mathsc{mpq}F_\mathsc{n}^{\,\,\,\mathsc{pq}} + g_\mathsc{mn} \frac{1}{6} F_\mathsc{pqr}F^\mathsc{pqr} \right)+\frac{1}{4\times 4!}e^{2\Phi}F_\mathsc{mpqrs}F_\mathsc{n}^{\,\,\,\mathsc{pqrs}},
\end{equation}
for the metric.

\bibliographystyle{JHEP}

\bibliography{Stijnsbibfile}

\end{document}